\documentclass[iop,apj]{emulateapj}

\begin{document}
\newcommand{\gps}{\ensuremath{g_{\rm P1}}}
\newcommand{\rps}{\ensuremath{r_{\rm P1}}}
\newcommand{\ips}{\ensuremath{i_{\rm P1}}}
\newcommand{\zps}{\ensuremath{z_{\rm P1}}}
\newcommand{\yps}{\ensuremath{y_{\rm P1}}}
\newcommand{\wps}{\ensuremath{w_{\rm P1}}}
\newcommand{\grizy}{\gps\rps\ips\zps\yps}
\newcommand{\griz}{\gps\rps\ips\zps}
\newcommand{\PS}{\protect \hbox {Pan-STARRS1}}
\newcommand{\afx}{PS1-10afx}
\newcommand{\msun}{M$_{\sun}$}
\newcommand{\kms}{km~s$^{-1}$}
\newcommand{\synow}{\texttt{SYNOW}}
\newcommand{\dm}{$\Delta$$m_{15}$}
\newcommand{\tbb}{\ensuremath{T_{\mathrm{BB}}}}
\newcommand{\rbb}{\ensuremath{R_{\mathrm{BB}}}}
\newcommand{\lbol}{\ensuremath{L_{\mathrm{bol}}}}
\newcommand{\mbol}{\ensuremath{M_{\mathrm{bol}}}}

\shorttitle{Unique SLSN at $z$ = 1.4}
\shortauthors{Chornock et al.}

\title{PS1-10\lowercase{afx} at $z=1.388$: Pan-STARRS1 Discovery of a
  New Type of Superluminous Supernova} 

\author{R.~Chornock\altaffilmark{1},
E.~Berger\altaffilmark{1},
A.~Rest\altaffilmark{2},
D.~Milisavljevic\altaffilmark{1},
R.~Lunnan\altaffilmark{1},
R.~J.~Foley\altaffilmark{1,3},
A.~M.~Soderberg\altaffilmark{1},
S.~J.~Smartt\altaffilmark{4},
A.~J.~Burgasser\altaffilmark{5},
P.~Challis\altaffilmark{1},
L.~Chomiuk\altaffilmark{6},
I.~Czekala\altaffilmark{1},
M.~Drout\altaffilmark{1},
W.~Fong\altaffilmark{1},
M.~E.~Huber\altaffilmark{7},
R.~P.~Kirshner\altaffilmark{1},
C.~Leibler\altaffilmark{8},
B.~McLeod\altaffilmark{1},
G.~H.~Marion\altaffilmark{1},
G.~Narayan\altaffilmark{1},
A.~G.~Riess\altaffilmark{2,9},
K.~C.~Roth\altaffilmark{10},
N.~E.~Sanders\altaffilmark{1},
D.~Scolnic\altaffilmark{9},
K.~Smith\altaffilmark{3},
C.~W.~Stubbs\altaffilmark{1,11},
J.~L.~Tonry\altaffilmark{7},
S.~Valenti\altaffilmark{12},
W.~S.~Burgett\altaffilmark{7},
K.~C.~Chambers\altaffilmark{7},
K.~W.~Hodapp\altaffilmark{7},
N. Kaiser\altaffilmark{7},
R.-P.~Kudritzki\altaffilmark{7},
E.~A.~Magnier\altaffilmark{7},
and P.~A.~Price\altaffilmark{13}
}

\altaffiltext{1}{Harvard-Smithsonian Center for Astrophysics, 60
  Garden St., Cambridge, MA 02138, USA, \texttt{rchornock@cfa.harvard.edu}}
\altaffiltext{2}{Space Telescope Science Institute, 3700 San Martin
  Drive, Baltimore, MD 21218, USA} 
\altaffiltext{3}{Clay Fellow}
\altaffiltext{4}{Astrophysics Research Centre, School of Mathematics
  and Physics, Queen's University Belfast, Belfast, BT7 1NN, UK} 
\altaffiltext{5}{Center for Astrophysics and Space Science, University
  of California San Diego, La Jolla, CA 92093, USA}
\altaffiltext{6}{Jansky Fellow; Department of Physics and Astronomy,
  Michigan State University, East Lansing, MI 48824, USA} 
\altaffiltext{7}{Institute for Astronomy, University of Hawaii, 2680
  Woodlawn Drive, Honolulu HI 96822, USA} 
\altaffiltext{8}{Department of Astronomy \& Astrophysics, University of
  California, Santa Cruz, CA 95060, USA} 
\altaffiltext{9}{Department of Physics and Astronomy, Johns Hopkins
  University, 3400 North Charles Street, Baltimore, MD 21218, USA} 
\altaffiltext{10}{Gemini Observatory, 670 North Aohoku Place, Hilo, HI
  96720, USA}
\altaffiltext{11}{Department of Physics, Harvard University, 17 Oxford
  Street, Cambridge, MA 02138, USA}
\altaffiltext{12}{Las Cumbres Observatory Global Telescope Network,
  Inc., Santa Barbara, CA 93117, USA; Department of Physics,
  University of California Santa Barbara, Santa Barbara, CA
  93106-9530, USA}
\altaffiltext{13}{Department of Astrophysical Sciences, Princeton
  University, Princeton, NJ 08544, USA}

\begin{abstract}
We present the \PS\ discovery of \afx, a unique
hydrogen-deficient superluminous supernova (SLSN) at redshift
$z=1.388$.  The light curve peaked at $\zps=21.7$~mag, making
\afx\ comparable to the most luminous known SNe, with $M_u =
-22.3$~mag.  Our extensive optical and near-infrared observations 
indicate that the bolometric light
curve of \afx\ rose on the unusually fast timescale of $\sim$12~d to
the extraordinary peak luminosity of 4.1$\times$10$^{44}$ erg s$^{-1}$
($\mbol = -22.8$ mag) and subsequently faded rapidly.  Equally
important, the SED is unusually red for a SLSN, with a color
temperature of $\sim$6800 K near maximum light, in contrast to
previous hydrogen-poor SLSNe, which are bright in the ultraviolet (UV).
The spectra more closely resemble those of a normal  
SN Ic than any known SLSN, with a photospheric velocity of
$\sim$$11,000$~\kms\ and evidence for line blanketing in the
rest-frame UV.    Despite the fast rise, these parameters imply a very
large emitting radius ($\gtrsim$5$\times$10$^{15}$ cm). We
demonstrate that no existing theoretical 
model can satisfactorily explain this combination of properties:
(i) a nickel-powered light curve cannot match the combination of high peak
luminosity with the fast timescale; (ii) models powered by the spindown
energy of a rapidly-rotating magnetar predict significantly hotter and
faster ejecta; and (iii) models invoking shock breakout through a dense
circumstellar medium cannot explain the observed spectra or color
evolution.  
The host galaxy is well detected in pre-explosion imaging
with a luminosity near $L^{*}$, a star formation rate of $\sim$15
\msun yr$^{-1}$,
and is fairly massive ($\sim$2$\times$10$^{10}$~\msun), with a stellar
population age of $\sim$10$^8$~yr, also in contrast to the young dwarf
hosts of known hydrogen-poor SLSNe.   
\afx\ is distinct from known examples of SLSNe in its spectra, colors,
light-curve shape, and host galaxy properties, suggesting that it
resulted from a different channel than other hydrogen-poor SLSNe.
\end{abstract}
\keywords{supernovae: individual (PS1-10afx) --- surveys (Pan-STARRS1)}

\section{Introduction}

A small fraction of massive stars end their lives with spectacular
explosions one or two orders of magnitude more luminous than normal
supernovae (SNe).  After the initial puzzling discoveries of the
luminous SNe~2005ap \citep{quimby05ap} and 2006gy 
\citep{ofek06gy,smith06gya}, modern wide-field surveys over the past
decade began to uncover these superluminous SNe (SLSNe) in greater
numbers. The energy 
scales involved in these  explosions challenge our understanding of
conventional SN explosions. Normal SNe resulting from iron core
collapse have characteristic energy scales of $\sim$10$^{51}$ erg of
kinetic energy and $\sim$10$^{49}$ erg emitted
as optical radiation ($\sim$10$^{42}$ erg s$^{-1}$ for $\sim$10$^7$
s).  The SLSNe are far off this scale--- they peak at optical
luminosities of up to $\sim$4$\times 10^{44}$ erg s$^{-1}$
\citep{quimby11,laura,miller08es,gezari08es} and emit
a total of up to 4$\times$10$^{51}$ erg optically \citep{rest03ma}.

This large energy scale motivates the question of what physics powers
these SNe, and how to accommodate these objects within
the standard understanding of massive star evolution.  Theorists
have proposed a number of exotic power 
sources, including the pair instability mechanism (e.g.,
\citealt{barkat67,rakavy67}) and reprocessed spindown energy released
by a newly formed magnetar \citep{woosley10,kb10}.  Another
possibility is interaction with a dense circumstellar medium (CSM)
\citep{ci11,mt12,gb12,manos12}, requiring extreme CSM masses and
densities whose origin remains unexplained (see \citealt{woosley07}
for one possibility).  All of these models require additional
ingredients beyond the normal stellar evolutionary processes.

\citet{galyam12} has attempted to impose order on the menagerie of
objects achieving sufficient peak luminosities to be
classified as SLSNe ($M < -21$ mag is a typical requirement) by
sorting them into three categories.  All of the hydrogen-rich objects
were classified as SLSNe-II and all exhibit signs of being powered by
dense CSM interaction, with the possible exception of SN~2008es
\citep{miller08es,gezari08es}.   He split the objects lacking hydrogen
into two 
classes, the rare SLSNe-R that have slow photometric decline rates
consistent with being powered by the radioactive decay of a very large
synthesized mass of $^{56}$Ni, and the relatively homogeneous class of
SLSNe-I, whose power source is still mysterious.  A few caveats have
been raised.  The SLSNe-R are interpreted to be the results of
pair-instability SNe.  However, existing models for the pair
instability process prefer extremely low
metallicity, and may be in conflict with the observed spectrum and
spectral energy distribution (SED) of SLSNe-R (e.g.,
\citealt{dessart_pi}).  Also, it is not clear how homogeneous the 
SLSNe-I class really is.  Although the spectra of most appear to be
similar to those of SN~2005ap and SCP06F6 \citep{quimby05ap,barbary},
the rise times and peak luminosities of published objects vary by
factors of $\sim$5
\citep{quimby05ap,10gx,quimby11,laura,06oz,bzj}. All SLSNe-I to date
have had hot spectra and been bright in the rest-frame
near-ultraviolet (NUV) relative to normal SN SEDs.

In this paper, we present the discovery of \afx, an extreme SLSN at
redshift $z=1.388$ that does not fit into this classification scheme
and is distinct from all previous SLSNe.
The peak luminosity is comparable to the highest known and 
the rise time is the fastest measured.  The spectra show no evidence
for hydrogen and lack any analog in the existing sample of SLSNe.
Instead, they most closely resemble those of line-blanketed
normal SNe Ic. 
In Section 2, we present the suite of optical and near-infrared (NIR)
observations.  The host galaxy is described in Section 3. We compare
our observations of \afx\ to known SNe in Section 4. 
In Section 5, we construct the SED and
bolometric light curve.  We 
then compare \afx\ to existing SLSN models in Section 6.
All calculations in this paper assume a flat $\Lambda$CDM cosmology
with $H_0$=74~\kms~Mpc$^{-1}$, $\Omega_m$=0.27, and
$\Omega_{\Lambda}$=0.73 \citep{WMAP,shoes}.

\section{Observations}

\subsection{\PS}

The Pan-STARRS1 (PS1) telescope has a
1.8~m diameter primary mirror that images a field with a diameter of
3.3\degr 
\citep{PS1_optics} onto a total of sixty $4800\times4800$ pixel
detectors, with a pixel scale of 0.258\arcsec \citep{PS1_GPCA}.
A more complete description of the PS1 system, hardware
and software, is provided by \cite{PS1_system}. 

The PS1 observations are obtained through a set of five broadband
filters, designated as \gps, \rps, \ips, \zps, and
\yps.   Although the filter system for PS1 has much in
common with that used in previous surveys, such as the Sloan Digital
Sky Survey (SDSS; \citealt{SDSS}), there
are differences.  Most important for this work, the 
\zps\ filter is cut off at 9300~\AA, giving it a different response than
the detector response defined $z_{SDSS}$, and SDSS has no corresponding
\yps\ filter.  Further information on the passband shapes is described
by \cite{PS_lasercal}.    Photometry is in the ``natural'' PS1 
system, $m=-2.5\log(f_{\nu})+m'$, with a single zeropoint adjustment $m'$
made in each band to conform to the AB magnitude scale
\citep{JTphoto}.  Photometry from all other sources presented in this
paper is also on the AB scale.  PS1 magnitudes are interpreted as
being at the top of the atmosphere, with 1.2 airmasses of atmospheric
attenuation being included in the system response function.

The PS1 Medium Deep Survey (MDS) consists of 10 fields across the sky
that are observed nightly when in season ($\sim$5 months per year)
with a typical cadence of 3~d between observations in \griz\ in dark
and gray time, while \yps\ is used near full moon.  PS1 data are
processed 
through the Image Processing Pipeline (IPP; \citealt{PS1_IPP}) on a
computer cluster at the Maui High Performance Computer Center. The
pipeline runs the images through a succession of stages,  
including flat-fielding (``de-trending''), a flux-conserving warping
to a sky-based image plane, masking and artifact removal, and object
detection and photometry.  Transient detection using IPP photometry is
carried out at Queen’s University Belfast. Independently, difference
images are produced from the stacked nightly MDS images by the
\texttt{photpipe} pipeline \citep{rest05} running on the Odyssey
computer cluster at Harvard University. The discovery and data
presented here are from the \texttt{photpipe} analysis.

\begin{figure*}
\epsscale{1.2}
\plotone{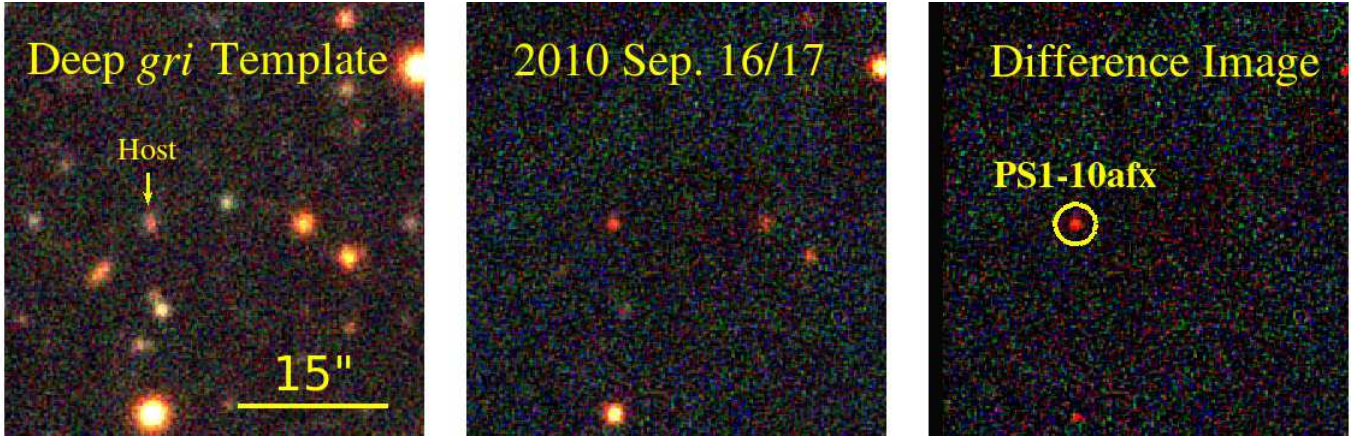}
\caption{Three-color \gps\rps\ips\ images of the field of \afx, showing
  ({\it left}) a deep stack of pre-explosion imaging with the host
  galaxy marked; ({\it center}) images taken near maximum light; and
  ({\it right}) difference images of the templates subtracted from the
  observations.  The color scales are similar in each panel.  The
  unusually red color of \afx\ compared to other faint objects in the
  field is apparent, as it is only
  strongly detected in \ips, with non-detections in \gps\ and \rps.
}
\label{hostfig}
\end{figure*}

\afx\ was first detected in MDS imaging on 2010 August 31.35 (UT
dates are used throughout this paper) at a position of 
$\alpha$ = 22$^{\mathrm  h}$11$^{\mathrm m}$24\fs162, 
$\delta$ = $+$00\degr09$'$43\farcs49 (J2000), with an uncertainty of
0$\farcs$1 in each coordinate.  The strong detections in \ips\ and
\zps\ combined with non-detections in \gps\ and \rps\ over the next
few nights immediately garnered our attention. The unusual colors are
evident in the color images of the field presented in
Figure~\ref{hostfig}. 

\begin{figure}
\epsscale{1.2}
\plotone{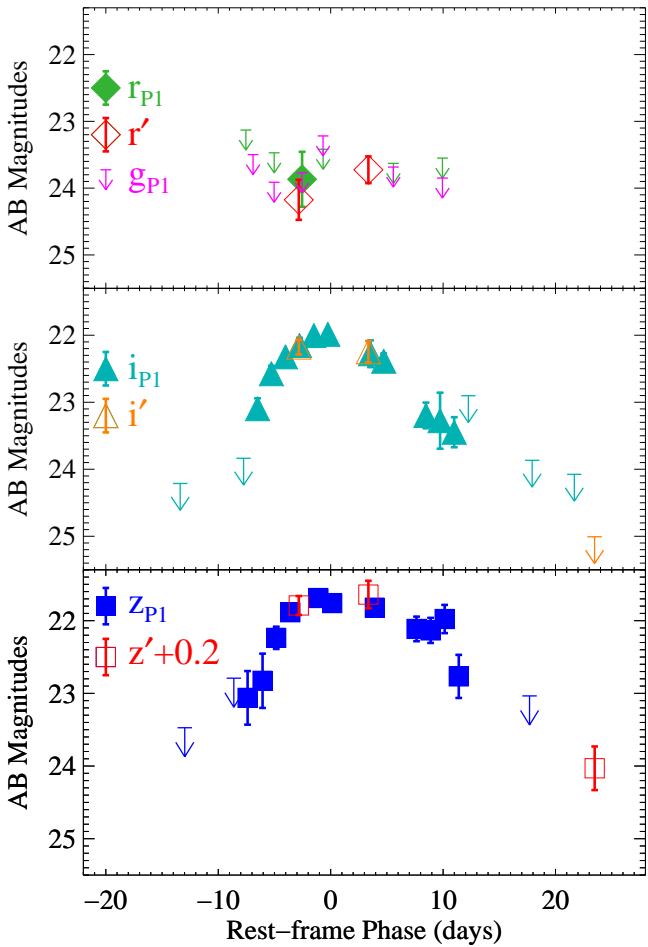}
\caption{Multicolor optical light curves of \afx.  Most of the data
  were obtained with PS1, but points marked $r'$, $i'$, or $z'$ come
  from other telescopes (see Table~\ref{phottab}).  The $r'i'z'$
  photometry points near days $-$3 and $+$3 were taken
  contemporaneously with the optical spectra shown in
  Figure~\ref{specplot}. Upper limits marked with arrows correspond to
  3$\sigma$. 
}
\label{multiplot}
\end{figure}

We constructed deep template stacks from all pre-explosion images and
subtracted them from the PS1 observations using \texttt{photpipe}
\citep{rest05}.  Details of the photometry and generation of PS1 SN
light curves will be given by Rest et al. (2013, in prep.) and Scolnic
et al. (2013, in prep.). 
The typical spacing between MDS observations in the same filter 
corresponds to only 1.3~d in the rest frame of \afx, so in some cases
we have co-added the photometry from adjacent observations to
increase the significance of marginal detections or the depth of the
non-detections.  The final \afx\ photometry, after correction for
$E(B-V) = 0.05$~mag of Galactic extinction \citep{eddiedoug,sfd98}, is
given in Table~\ref{phottab} and shown in Figure~\ref{multiplot}.  

We fit polynomials to the \zps\ data points within 15
rest-frame days of peak to determine the time of maximum light, for
which we adopt a Modified Julian Date (MJD) of 55457.0 (=2010
September 18.0).  All phases referred to subsequently are in rest-frame
days referenced from this date. 

\subsection{Other Photometry}

In addition to the PS1 observations, we obtained two epochs of
multicolor photometry using the Gemini Multi-Object Spectrographs
(GMOS; \citealt{hookgmos}) on the 8-m Gemini-North and South telescopes
and one epoch of imaging using the Inamori-Magellan Areal Camera and
Spectrograph 
(IMACS; \citealt{imacspaper}) on the 6.5-m Magellan Baade telescope.
The images were processed using standard 
tasks and then archival fringe frames were subtracted from the GMOS
images using the \texttt{gemini} IRAF\footnote{IRAF is
  distributed by the National Optical Astronomy Observatories, 
    which are operated by the Association of Universities for Research
    in Astronomy, Inc., under cooperative agreement with the National
    Science Foundation.} package.  We used several SDSS stars in the
field to calibrate the Gemini images, while the IMACS zeropoints
were checked with observations of standard star fields obtained the
same night.  We subtracted the deep PS1 templates in the corresponding
filter from each image using the ISIS software package
\citep{isis} to correct for host 
contamination.  The final photometry is listed in Table~\ref{phottab}.
The $r'$ and $i'$ magnitudes agree well with the PS1 observations at
similar epochs.  However, the $z'$ observations exhibit an offset
from the \zps\ light curve due to the filter response differences
noted above (GMOS and IMACS are closer to SDSS).  We therefore
add 0.2 mag to the $z'$ photometry points for consistency with
\zps\ whenever we refer to the combined $z$-band light 
curve. Finally, we obtained late-time observations of the host galaxy
in $g'$ and $r'$ on 2011 October 21.1 (phase $+$167~d) using 
the Low Dispersion Survey Spectrograph-3 (LDSS3) on the 6.5-m
Magellan Clay telescope (Table~\ref{hosttab}). 

\begin{deluxetable}{lrcccc}
\tabletypesize{\scriptsize}
\tablecaption{\afx\ Photometry}
\tablehead{
\colhead{MJD} &
\colhead{Epoch\tablenotemark{a}} &
\colhead{Filter} &
\colhead{Magnitude\tablenotemark{b}} &
\colhead{Error} & 
\colhead{Instrument} \\
 & \colhead{(d)} & & (AB) & &
}
\startdata
55440.5\tablenotemark{c} &  $-$6.9 & \gps & $>$23.50 & \nodata & PS1 \\
55445.0\tablenotemark{c} &  $-$5.0 & \gps & $>$23.91 & \nodata & PS1 \\
55450.9\tablenotemark{c} &  $-$2.6 & \gps & $>$23.77 & \nodata & PS1 \\
55455.4\tablenotemark{c} &  $-$0.7 & \gps & $>$23.22 & \nodata & PS1 \\
55470.3\tablenotemark{c} &   5.6 & \gps & $>$23.68 & \nodata & PS1 \\
55480.7\tablenotemark{c} &   9.9 & \gps & $>$23.85 & \nodata & PS1 \\
\hline
55439.0\tablenotemark{c} &  $-$7.5 & \rps & $>$23.13 & \nodata & PS1 \\
55445.0\tablenotemark{c} &  $-$5.0 & \rps & $>$23.47 & \nodata & PS1 \\
55450.2 &  $-$2.8 & $r'$ & 24.18 & 0.30 & GMOS-N \\
55450.9\tablenotemark{c} &  $-$2.5 & \rps & 23.87 & 0.41 & PS1 \\
55455.4\tablenotemark{c} &  $-$0.7 & \rps & $>$23.42 & \nodata & PS1 \\
55465.0 &   3.4 & $r'$ & 23.73 & 0.20 & GMOS-S \\
55470.3\tablenotemark{c} &   5.6 & \rps & $>$23.63 & \nodata & PS1 \\
55480.8\tablenotemark{c} &   9.9 & \rps & $>$23.55 & \nodata & PS1 \\
\hline
55425.0\tablenotemark{c} & $-$13.4 & \ips & $>$24.21 & \nodata & PS1 \\
55438.5\tablenotemark{c} &  $-$7.7 & \ips & $>$23.84 & \nodata & PS1 \\
55441.5 &  $-$6.5 & \ips & 23.10 & 0.16 & PS1 \\
55444.4 &  $-$5.3 & \ips & 22.57 & 0.12 & PS1 \\
55447.4 &  $-$4.0 & \ips & 22.32 & 0.06 & PS1 \\
55450.4 &  $-$2.8 & \ips & 22.15 & 0.11 & PS1 \\
55450.2 &  $-$2.8 & $i'$ & 22.18 & 0.11 & GMOS-N \\
55453.4 &  $-$1.5 & \ips & 22.00 & 0.05 & PS1 \\
55456.4 &  $-$0.3 & \ips & 21.99 & 0.05 & PS1 \\
55465.0 &   3.4 & $i'$ & 22.25 & 0.16 & GMOS-S \\
55465.5 &   3.5 & \ips & 22.28 & 0.20 & PS1 \\
55468.3 &   4.7 & \ips & 22.39 & 0.12 & PS1 \\
55477.2 &   8.5 & \ips & 23.20 & 0.19 & PS1 \\
55480.2 &   9.7 & \ips & 23.28 & 0.42 & PS1 \\
55483.2 &  11.0 & \ips & 23.45 & 0.22 & PS1 \\
55486.3 &  12.2 & \ips & $>$22.90 & \nodata & PS1 \\
55499.8\tablenotemark{c} &  17.9 & \ips & $>$23.87 & \nodata & PS1 \\
55508.8\tablenotemark{c} &  21.7 & \ips & $>$24.08 & \nodata & PS1 \\
55513.1 &  23.5 & $i'$ & $>$25.01 & \nodata & IMACS \\
\hline
55426.0\tablenotemark{c} & $-$13.0 & \zps & $>$23.47 & \nodata & PS1 \\
55436.4 &  $-$8.6 & \zps & $>$22.79 & \nodata & PS1 \\
55439.3 &  $-$7.4 & \zps & 23.06 & 0.37 & PS1 \\
55442.5 &  $-$6.1 & \zps & 22.83 & 0.37 & PS1 \\
55445.4 &  $-$4.8 & \zps & 22.24 & 0.15 & PS1 \\
55448.4 &  $-$3.6 & \zps & 21.88 & 0.12 & PS1 \\
55450.2 &  $-$2.8 & $z'$ & 21.59 & 0.13 & GMOS-N \\
55454.4 &  $-$1.1 & \zps & 21.69 & 0.10 & PS1 \\
55457.3 &   0.1 & \zps & 21.76 & 0.11 & PS1 \\
55465.0 &   3.4 & $z'$ & 21.44 & 0.19 & GMOS-S \\
55466.4 &   3.9 & \zps & 21.82 & 0.10 & PS1 \\
55475.2 &   7.6 & \zps & 22.11 & 0.17 & PS1 \\
55478.2 &   8.9 & \zps & 22.13 & 0.17 & PS1 \\
55481.2 &  10.1 & \zps & 21.98 & 0.19 & PS1 \\
55484.2 &  11.4 & \zps & 22.77 & 0.30 & PS1 \\
55499.2\tablenotemark{c} &  17.7 & \zps & $>$23.04 & \nodata & PS1 \\
55513.1 &  23.5 & $z'$ & 23.83 & 0.30 & IMACS \\
\hline
55463.4 &   2.7 & \yps & 21.09 & 0.16 & PS1 \\
55492.8\tablenotemark{c} &  15.0 & \yps & $>$21.95 & \nodata & PS1 \\
55523.3 &  27.8 & \yps & $>$21.82 & \nodata & PS1 \\
55463.4 &   2.7 & $Y$ & 20.96 & 0.10 & NIRI \\
\hline
55463.3 &   2.6 & $J$ & 20.99 & 0.17 & MMIRS \\
55463.4 &   2.7 & $J$ & 21.19 & 0.10 & NIRI \\
55481.1 &  10.1 & $J$ & 22.22 & 0.26 &  MMIRS \\
55485.0 &  11.7 & $J$ & 22.28 & 0.30 &  MMIRS \\
55496.0 &  16.3 & $J$ & 22.19 & 0.10 &  HAWK-I \\
55515.1 &  24.3 & $J$ & 22.84 & 0.14 &  HAWK-I \\
55552.0 &  39.8 & $J$ & $>$21.90 & \nodata & HAWK-I \\
55561.0 &  43.6 & $J$ & $>$21.90 & \nodata & HAWK-I \\
\hline
55463.4 &   2.7 & $H$ & 21.20 & 0.19 & NIRI \\
55485.0 &  11.7 & $H$ & 21.05 & 0.44 & MMIRS \\
55490.1 &  13.9 & $H$ & 21.94 & 0.24 & HAWK-I \\
55515.1 &  24.3 & $H$ & 22.28 & 0.39 & HAWK-I \\
\hline
55463.4 &   2.7 & $K$ & 21.46 & 0.26 & NIRI \\
55485.0 &  11.7 & $K_s$ & $>$21.35 & \nodata & MMIRS
\enddata
\tablenotetext{a}{In rest-frame days relative to maximum light on MJD
  55457.0. }
\tablenotetext{b}{Corrected for Galactic reddening.  Upper limits are 3$\sigma$.}
\tablenotetext{c}{Data point is from a stack of observations taken on
  multiple nights.  MJD reported is mean of individual observations in
  stack.}
\label{phottab}
\end{deluxetable}

The high redshift and red optical colors of \afx\ indicate that NIR
photometry is required to constrain the SED and bolometric luminosity.
We received director's discretionary (DD) time at Gemini-North to
obtain $YJHK$ photometry near maximum light using the Near InfraRed
Imager and Spectrometer (NIRI; \citealt{niri}).  The observations were
acquired on 2010 September 24.4, at a phase of $+$2.7~d.  Additional
observations were obtained over the subsequent month in $JHK_{s}$ 
with the MMT \& Magellan Infrared Spectrograph (MMIRS;
\citealt{mmirs}) on the Magellan Clay Telescope. To follow the 
light curve in the rest-frame optical to late times, we obtained a
series of $JH$ observations using DD time on the 8-m Very Large Telescope
(Yepun) with the High-Acuity Wide-field K-band Imager (HAWK-I;
\citealt{hawki}). Finally, we used the FourStar Infrared Camera
\citep{fourstar} on the Magellan Baade telescope to obtain
late-time $JHK_{s}$ observations of the host galaxy of \afx\ on 2011
September 18, 2011 September 21, and 2011 December 7 
(phases $+$153~d, $+$154~d, and $+$186~d),
respectively (Table~\ref{hosttab}).

\begin{figure}
\epsscale{1.2}
\plotone{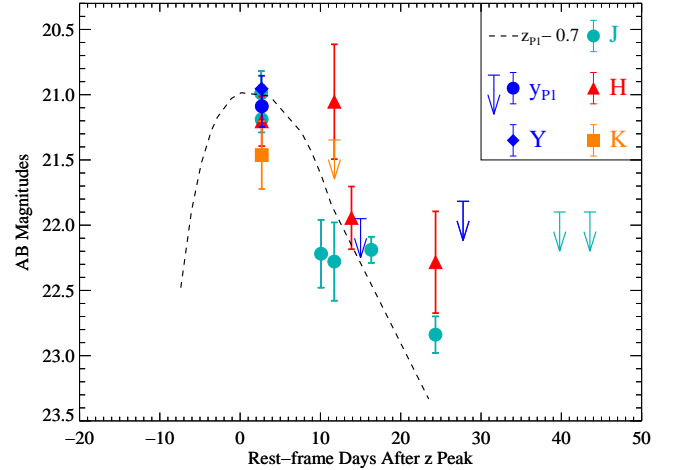}
\caption{NIR photometry of \afx.  The dashed line shows a polynomial
  fit to the well-sampled 
  \zps\ light curve to guide the eye, shifted by 0.7 mag to go through
  the $J$ photometry near maximum light.
}
\label{nirplot}
\end{figure}

The images were flat fielded, sky subtracted, and stacked using
standard tasks in IRAF, including the \texttt{gemini} package for the 
NIRI data, except for the HAWK-I data, which were processed using the
instrument pipeline.  The images from each instrument were then
calibrated using the same 2MASS stars in each field,
except for the $Y$ image, which was calibrated using archival NIRI
zeropoints.  The $H$ and $K_s$ host galaxy fluxes were then subtracted
numerically from each datapoint in those filters.  For the $Y$ data
point, we interpolated the host galaxy flux between the measured
values in \yps\ and $J$ to
subtract off the host galaxy contribution (only a small correction).
We were able to perform image subtraction on most of the $J$ images
using the ISIS software package with the late-time FourStar image as a
template, but the others had the host $J$ flux subtracted numerically.
The final host-corrected photometry points are shown in
Figure~\ref{nirplot} and listed in Table~\ref{phottab}, including the
calibration uncertainty in the errors.

\subsection{Spectroscopy}

After noticing the unusually red $\rps - \ips$ colors of \afx, we
immediately triggered optical spectroscopy using GMOS-N.  A pair of
dithered 1800~s observations 
were taken on 2010 September 11.26 (phase $-2.8$~d) with the R400
grating and OG515 order-blocking filter to cover the wavelength range
of 5400$-$9650~\AA.  We used standard tasks in IRAF to perform basic
two-dimensional image processing and spectral extraction.  We
used our own IDL procedures to apply a flux calibration and correct
for telluric absorption based on archival observations of
spectrophotometric standard stars.  The observations were performed at
a mean airmass of 1.5, but the 1\arcsec-wide long slit was oriented
within 20\degr\ of the parallactic angle \citep{fil82}, so the overall
spectral shape is reliable at these red wavelengths.

A second epoch of spectroscopy was obtained on 2010 September 26.04
(phase $+$3.4~d) using GMOS-S.  The
observations and instrumental setup were similar to the first epoch,
except that a redder grating tilt was used to cover the range
5880$-$10100~\AA.  This spectrum has a lower signal-to-noise ratio
(S/N) than the first one, but the main SN features are still present.
We attempted a final epoch of spectroscopy using GMOS-S in
nod-and-shuffle mode on 2010 November 5, but \afx\ had faded
significantly, so we only detected a faint continuum with a
hint of SN features.  We do not consider this spectrum further.  

\begin{figure*}
\plotone{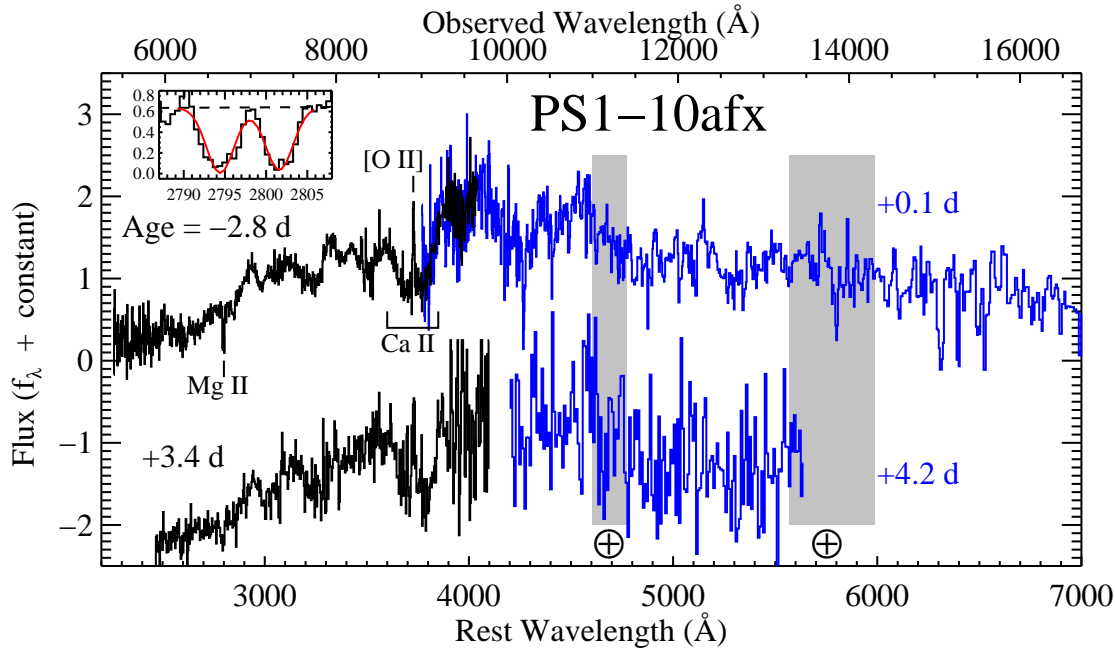}
\caption{Optical (black) and NIR (blue) spectra of \afx.  The spectra
  are labeled with their phase in rest-frame days.  The \ion{O}{2}
  $\lambda$3727 emission and Mg II $\lambda$2800 absorption from the
  host galaxy are labeled, as is the broad Ca II H+K P-Cygni feature
  from the SN itself.  The zeropoint of the
  flux scale is appropriate for the day $-$2.8/$+$0.1 GMOS/FIRE
  spectra.  The day $+$3.4/$+$4.2 GMOS/MMIRS spectra have been shifted
  down by 2.5 flux units.  The gray bars labeled with a $\earth$
  symbol represent the regions of strong telluric absorption between
  $Y$/$J$ and $J$/$H$.  The inset shows the fit to Mg II $\lambda$2800
  absorption doublet.
}
\label{specplot}
\end{figure*}

We observed \afx\ with the Folded-port InfraRed Echellette (FIRE;
\citealt{fireref}) spectrograph on the Magellan Baade telescope on
2010 September 18.17 (phase $+$0.1~d), using the
high throughput, low-resolution prism mode to cover the range of
0.8$-$2.5~$\mu$m.  The spectral resolution is a strongly decreasing
function of wavelength, but is $R\approx500$ in $J$.  Six 150~s
dithered exposures were obtained for a total of 900~s of on-source
time.  Despite the short exposure time, the continuum of \afx\ is well
detected blueward of $\sim$1.6~$\mu$m.  We reduced and combined the
spectra using the FIREHOSE package, including a correction for 
telluric absorption obtained from observations of the A0V star HD
208368 using the algorithm of \citet{vacca03}.

We obtained a MMIRS spectrum on 2010 September 28 (phase $+$4.2~d), 
using the $zJ$
filter and $J$ grism to cover the range 1.0$-$1.34~$\mu$m.  A total of
90 minutes were spent on source in a series of dithered 300~s
exposures.  We used standard tasks in IRAF to combine and extract the
spectra. 

The final optical and NIR spectra are shown in Figure~\ref{specplot}.  All of our
optical spectra exhibit an emission line near 8903~\AA, which we
identify as [\ion{O}{2}] $\lambda$3727 emission from the host galaxy,
as well as absorption from the \ion{Mg}{2} $\lambda$2800 doublet at
the same redshift of $z=1.388$, which we adopt as the SN redshift.

\subsection{Radio Observations}

Popular models for SLSNe do not predict detectable radio emission
at high redshift, as discussed by \citet{laura}.  However, given its
unusual properties, we observed \afx\ on 2010 September 12.2 (phase
$-$2.4~d) with the Karl G. Jansky Very Large 
Array \citep{evla} as part of our NRAO Key Science Project ``Exotic
Explosions, Eruptions, and Disruptions: A New Transient Phase-Space,''
with 256~MHz of bandwidth centered at 4.96~GHz.  The data were reduced
with standard tasks in the Astronomical Image Processing System (AIPS;
\citealt{aips}), using J2212+0152 as the gain calibrator and 3C48 as
the absolute flux density calibrator. These observations yielded a
non-detection of 19$\pm$18~$\mu$Jy.

\section{Host Galaxy Observations}

We measure a host redshift of $z=1.3883\pm0.0001$ from 
fits to the [\ion{O}{2}] line in our highest S/N spectrum (phase
$-$2.8~d).  A double-Gaussian fit to the \ion{Mg}{2} $\lambda$2800
doublet absorption in the same
spectrum (inset in Figure~\ref{specplot}) is blueshifted by
120$\pm$12~\kms\ relative to the [\ion{O}{2}] rest frame.  Such an
offset is typical of rest-frame UV selected star-forming galaxies at
this redshift \citep{erb12} and was also seen in observations of the
SLSN PS1-11bam \citep{11bam}.  This effect has been interpreted to be
caused by absorption occurring in galactic-scale outflows driven by
star formation.   The rest-frame equivalent widths ($W_r$) of the two 
lines are $W_r$($\lambda$2796)=1.8$\pm$0.2~\AA\ and
$W_r$($\lambda$2803)=1.6$\pm$0.2~\AA. 
These values are larger than for PS1-11bam \citep{11bam}, but
slightly lower than the median of the intrinsic absorbers from
gamma-ray burst (GRB) host galaxies \citep{fynbo09}.  In addition,
\ion{Mg}{1} $\lambda$2852 absorption is present, but
the UV spectral slope of \afx\ is so red that the S/N rapidly
decreases to the blue, making it hard to confirm the presence of other
expected strong interstellar absorption lines, such as \ion{Fe}{2}
$\lambda$2600. 

We obtained photometry of the host galaxy of \afx\ in eight filters
from $g'$ to $K_s$ (Table~\ref{hosttab}) to probe the host stellar
population from the rest-frame NUV ($\sim$2000~\AA) to the NIR
($\sim$0.9~$\mu$m). 
The host is well detected in our deep \ips\zps\yps\ PS1 template
images, which we supplemented with the observations from other
facilities described above.  Host photometry was performed with a
consistent aperture of radius 1.7\arcsec\ in all filters.

\begin{figure}
\epsscale{1.1}
\plotone{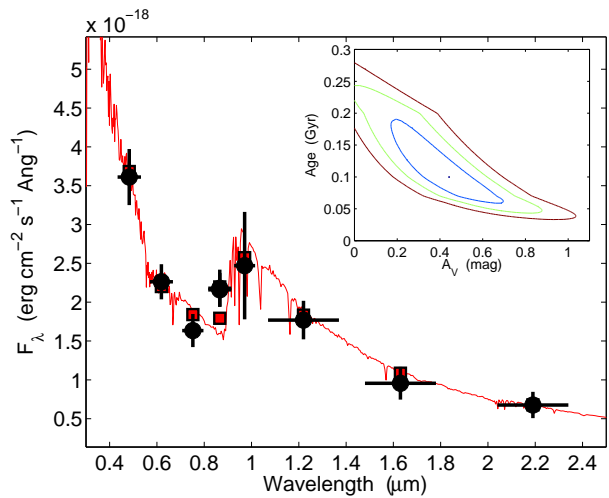}
\caption{Host galaxy SED.  The black data points represent the 
  observed host galaxy photometry (Table~\ref{hosttab}), corrected for
  the derived A$_V = 0.4$~mag of internal reddening, and plotted
  versus observed wavelength.  The filled red
  squares show the expected fluxes of the best-fit model (red line) in
  each of the observed bandpasses.  The inset shows the 1, 2, and
  3$\sigma$ contours for the best fit in the host reddening-age plane.
}
\label{hostplot}
\end{figure}

We fit a suite of single stellar population age models from
\citet{maraston05} to the 
data, assuming a red horizontal branch morphology, a Salpeter initial
mass function, and a metallicity of $Z=0.5~Z_{\odot}$.  The best-fit
model is shown in Figure~\ref{hostplot} and has an age of 10$^8$ yr and
requires a small amount of internal extinction (A$_V = 0.4$~mag).  The
derived stellar mass is $\sim$1.8 $\times$ 10$^{10}$~M$_{\odot}$.  The
$g'$ magnitude of the host galaxy corresponds to a NUV continuum
luminosity of 1.3$\times$10$^{29}$~erg~s$^{-1}$ (after correction for
internal extinction), which implies a star formation rate (SFR) of
$\sim$18 M$_{\odot}$~yr$^{-1}$ using the calibration of
\citet{kennicutt98}.  We also estimate a consistent value
of $\sim$13 \msun~yr$^{-1}$ from the observed [\ion{O}{2}] flux of 
$(4.7\pm0.5)\times$10$^{-17}$ erg~cm$^{-2}$~s$^{-1}$
\citep{kennicutt98}.  These SFRs would
be a factor of $\sim$2 lower without any correction for internal
extinction.

\begin{deluxetable}{ccccc}
\tabletypesize{\scriptsize}
\tablecaption{\afx\ Host Galaxy Photometry}
\tablehead{
\colhead{Filter} &
\colhead{Observed/Rest-frame} &
\colhead{Magnitude\tablenotemark{a}} &
\colhead{Error} & 
\colhead{Instrument} \\
 & \colhead{Wavelength (\AA)} & (AB) & &
}
\startdata
$g'$ & 4825/2020 & 24.01 & 0.10 & LDSS3 \\
$r'$ & 6170/2580 & 23.69 & 0.10 & LDSS3 \\
\ips & 7520/3150 & 23.43 & 0.13 & PS1 \\
\zps & 8660/3625 & 22.75 & 0.10 & PS1 \\
\yps & 9620/4030 & 22.29 & 0.28 & PS1 \\
$J$ & 12500/5230 & 22.02 & 0.14 & FourStar \\
$H$ & 16500/6910 & 21.93 & 0.22 & FourStar \\
$K_s$ & 21490/9000 & 21.53 & 0.25 & FourStar
\enddata
\tablenotetext{a}{Corrected for Galactic extinction.}
\label{hosttab}
\end{deluxetable}

This galaxy is significantly more massive than the previous hosts
of SLSNe.  \citet{neill11} examined the host galaxies of a sample
of luminous SNe and found them to have generally low stellar masses
and high specific SFRs.  In particular, all of the SLSNe-I in their
sample had dwarf hosts, with a median stellar mass of
$\sim$2$\times$10$^{8}$ \msun.  By comparison, \afx\ has a
significantly more massive host  with a lower specific SFR
($\sim$8$\times$10$^{-10}$~yr$^{-1}$) than their sample (with a median
of 3$\times$10$^{-9}$~yr$^{-1}$).  Only the luminous host galaxy of
the SLSN-II 2006gy exceeded our estimate for the stellar mass 
of the host of \afx.   Objects discovered subsequently have continued
this trend, having either dwarf host galaxies or no host at all
detected to date despite deep observations
\citep{quimby11,laura,06oz,chen,bzj}. 

While it has been argued that
the association of SLSNe with dwarf galaxy hosts is driven by their
low metallicity (e.g., \citealt{neill11}), the massive host of
\afx\ appears to be inconsistent with a low metallicity.  For example,
the relationships of \citet{fmr} imply a super-solar metallicity of
$12 + \log$(O/H) = 8.85 for our derived stellar mass and SFR, although
we caution that we cannot measure the metallicity directly with the
available data.  
An additional point of contrast is provided by the sample of
long-duration GRBs, whose occurrence is also widely believed to be
associated with low-metallicity environments (e.g.,
\citealt{stanek06}).  The host galaxy of \afx\ is brighter in $K_s$
than any of the GRB host galaxies at similar redshifts in the
optically-unbiased sample of \citet{hjorth12}, although a few of the
``dark'' GRB hosts studied by \citet{perley} are similarly luminous. 

The absolute magnitudes of \afx's host prior to any internal
extinction correction, $M_{2800{\mathrm \AA}} \approx -20.3$~mag and
$M_{B} \approx -21.7$~mag, are $\sim$0.4~mag brighter 
than $M^*$, the characteristic magnitudes in \citet{schechter}
function fits to the field galaxy luminosity functions at this
redshift \citep{gabasch04,faber07}.  The $K_s$ magnitude
of the host is also near the median of the distribution of
field galaxies at this redshift weighted by star formation rate
\citep{moircs}. Combined, these points suggest
that the environment of \afx\ is representative of typical
star-forming galaxies at this redshift, while being distinct from the
other hydrogen-poor SLSNe.

The host is marginally resolved in our template images, with a full
width at half-maximum (FWHM) of $\sim$1$\farcs$2, or 10 kpc at this
redshift.   Astrometric alignment of difference images taken near
maximum light with the templates shows that \afx\ is aligned with the
centroid of the host to within 0$\farcs$1 (0.8~kpc).  The other SLSN 
with a massive host galaxy, SN 2006gy \citep{ofek06gy,smith06gya}, is
also close to the nucleus, perhaps implying an unusual star
formation environment.

\section{Spectroscopic Comparisons}

\begin{figure}
\epsscale{1.15}
\plotone{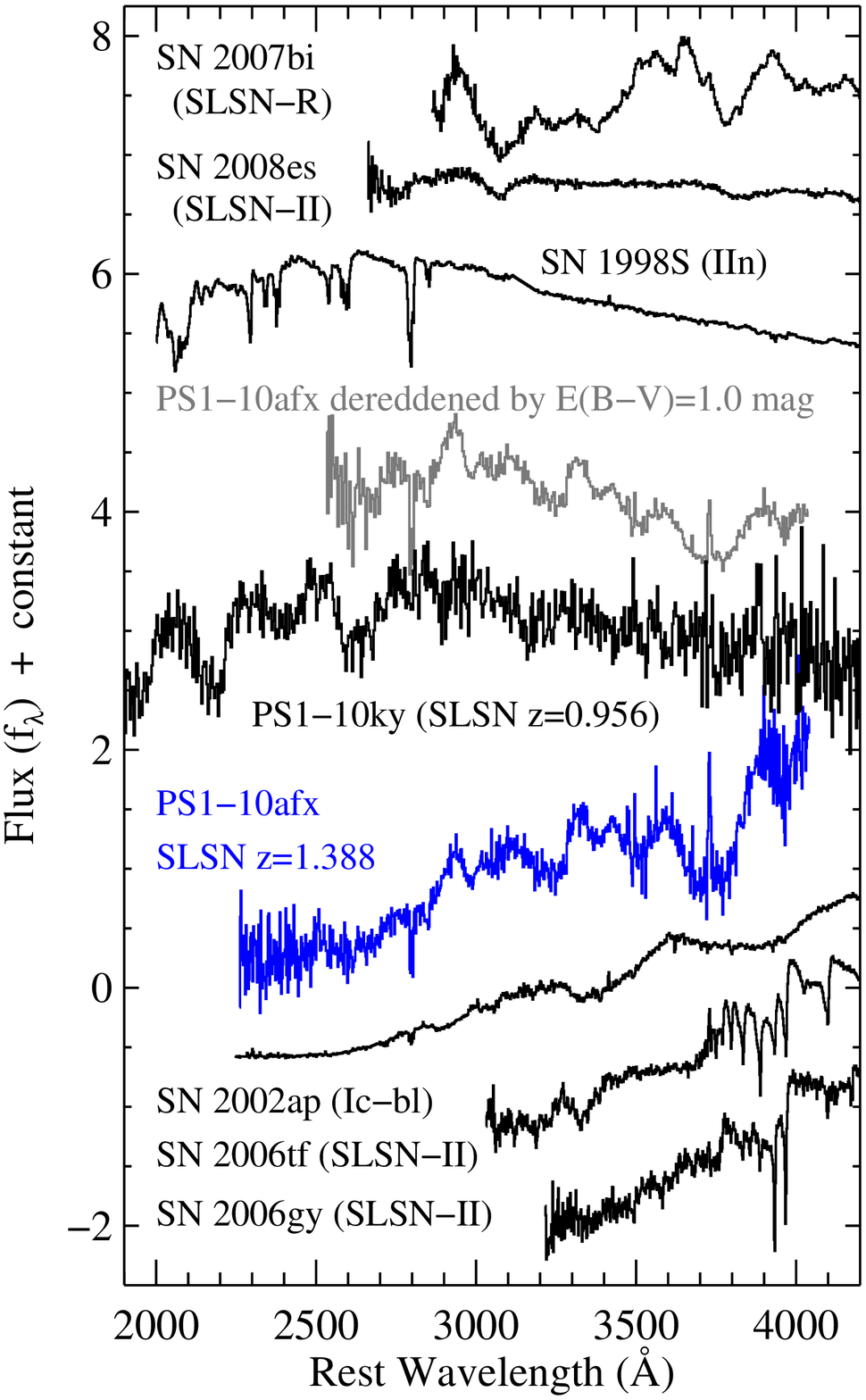}
\caption{SLSNe and other NUV spectral comparisons for \afx\ (blue;
  $-$2.8~d).   None of these objects is a good match to \afx.
  The spectrum in gray is the same one of \afx\ after correction
  for $E(B-V)=1.0$~mag of reddening (see text for details).  The
  other spectra and their phases relative to maximum light are: SN
  2007bi ($+$54~d; \citealt{galyam09}), SN
  2008es ($+$21~d; \citealt{miller08es}), SN 1998S ($+$14~d;
  \citealt{fransson98s}), PS1-10ky ($-$2~d; 
  \citealt{laura}), SN 2002ap ($+$4~d; data retrieved from the {\em
    Hubble Space Telescope} [{\em HST}]
  archive), SN 2006tf ($+$64~d; \citealt{smith06tf}), and SN 2006gy
  ($+$23~d; \citealt{smith06gyb}). All literature
  data are corrected for Galactic and host reddening, if known. 
}
\label{uvcomp}
\end{figure}

\begin{figure}
\epsscale{1.15}
\plotone{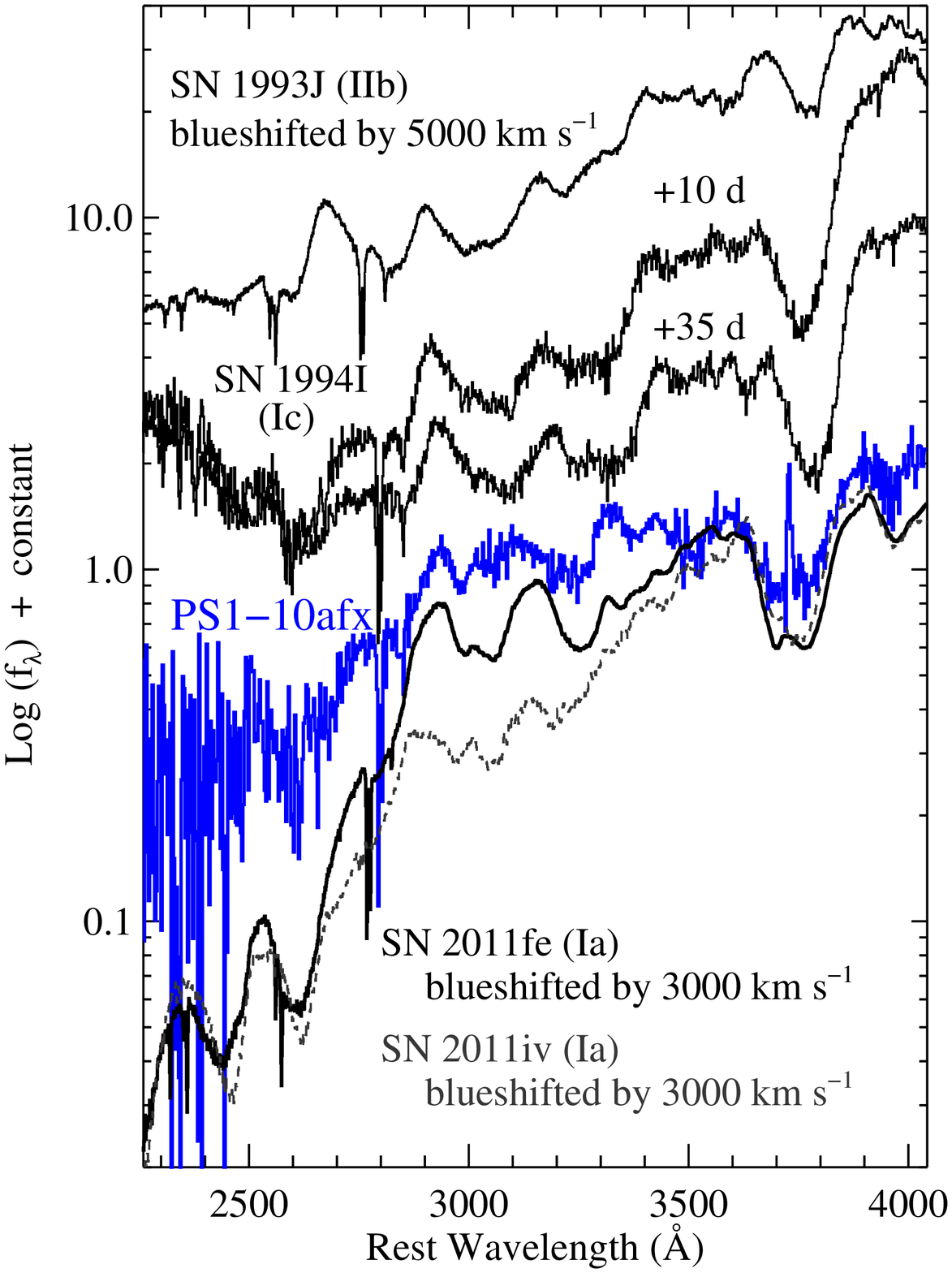}
\caption{Normal SNe NUV spectral comparisons for \afx\ (blue;
  $-$2.8~d).   
  These objects are much better matches than the ones shown in
  Figure~\ref{uvcomp}. The other spectra and their phases relative to
  maximum light are:  SN 1993J ($-$1~d; \citealt{jeffery93j}), SN
  1994I ($+$10~d and $+$35~d; data retrieved from the {\em HST}
  archive), and two SNe Ia, SN 2011fe ($+$0~d;
  \citealt{maguire11fe,foleyuv}) and SN 
  2011iv ($+$0.6~d; \citealt{foley11iv}).  The SNe Ia and SN 1993J
  spectra have been blueshifted to approximately match the absorption
  minima of the \ion{Ca}{2} feature.  All literature
  data are corrected for Galactic and host reddening, if known.
}
\label{uvcomplog}
\end{figure}

Despite its extraordinary luminosity (see Section 5), \afx\ has
spectra that more closely resemble those of a normal SN Ic than any
known SLSN.  We show our highest S/N spectrum (phase $-$2.8~d) in
Figures~\ref{uvcomp} and \ref{uvcomplog} along with several
comparisons drawn from the literature.  Relatively few UV spectra of
core-collapse SNe exist, but the available data allow us to sample a
wide variety of phenomena.  The most prominent feature in
the \afx\ spectrum is a broad P-Cygni feature with an absorption
minimum near 3730~\AA, which we identify as the typical \ion{Ca}{2}
H\&K absorption seen in most types of SNe, blueshifted by about
$16,000$~\kms.  There are several other weaker features in the
spectrum in the range 3000$-$3500~\AA.  The $+$3.4~d GMOS spectrum
is noisier, but similar, with some evidence that the broad \ion{Ca}{2}
absorption minimum decreased in velocity, although the overlapping
[\ion{O}{2}] emission and strong night sky residuals make this
uncertain.

These spectra are in strong contrast to the SN 2005ap-like class of
SLSNe \citep{quimby11}.  PS1-10ky is a high-redshift example of this
class \citep{laura} and it has a much bluer NUV continuum with no
\ion{Ca}{2} feature (Figure~\ref{uvcomp}).  A reasonable concern is
that \afx\ only appears to have redder colors due to host-galaxy
extinction.  However, matching the UV spectral slope of \afx\ to
PS1-10ky requires $E(B-V)\approx1$~mag of extinction (gray
line in Figure~\ref{uvcomp}), assuming a Galactic reddening curve
\citep{ccm}.  This is unsatisfactory for two reasons.  First,
\afx\ is already more luminous in the NUV than the SN~2005ap-like 
objects (see below), so $A_u\approx4.7$~mag would imply an upward
correction of two orders of magnitude to an already extreme peak
luminosity.  Second, the 
strongest UV spectral features of the SN 2005ap-like objects are the
trio of features blueward of 3000~\AA\ first seen in SCP06F6
\citep{barbary}, with the continuum redward of that being largely
smooth except for \ion{O}{2} features \citep{quimby11}.  \afx\ 
has fundamentally different spectral features with no correspondence
in the PS1-10ky spectrum, so the simplest explanation is that it has
an intrinsically cooler photosphere.  

The SN IIn 1998S is shown in Figure~\ref{uvcomp} as a well-studied
example of a SN undergoing strong CSM interaction.  SN 1998S at this
phase showed no broad P-Cygni features in the UV, but had a blue
continuum with narrow absorption lines from interstellar and
circumstellar material \citep{fransson98s}.  The SLSNe 2006gy and
2006tf are 
higher-luminosity versions of SNe dominated by hydrogen-rich CSM
interaction \citep{ofek06gy,smith06gya,smith06tf,smith06gyb}.  These
objects have redder spectra than SN 1998S, but still exhibit Balmer
lines and deep narrow absorptions that have no analog in the
\afx\ spectrum.    The SLSN
2008es had almost featureless blue spectra near maximum light
\citep{miller08es,gezari08es}, but the plotted spectrum is from a
later epoch when Balmer lines were beginning to develop strength in
the rest-frame optical spectrum.

The NUV spectrum of the
broad-lined SN Ic 2002ap is included because the engine-driven SNe
associated with GRBs have similar spectra, even if 2002ap
itself was not unusually luminous \citep{mazzali02ap}.  SN 2002ap has
much broader and more blended features than \afx.  Thus, none of these
objects in Figure~\ref{uvcomp} resembles \afx.  The SLSN 2007bi has
been identified as a potential 
pair-instability SN \citep{galyam09,young10}.  It is more similar to
\afx, with a definite \ion{Ca}{2} feature, but is still significantly
bluer and has several additional spectral features that do not match.

Instead, the set of spectra from normal SNe plotted in
Figure~\ref{uvcomplog} provide a much better basis for comparison. 
The SNe Ia and SN 1993J (Type IIb)
spectra have been blueshifted by the indicated amounts
to approximately match the \ion{Ca}{2} absorption minima to \afx.
Over the range 2800$-$4000~\AA, the spectra in Figure~\ref{uvcomplog}
are all very similar, with the dominant features being due to
\ion{Ca}{2} in all objects.  A notch near 3950~\AA\ is from
\ion{Si}{2} $\lambda$4130.  An emission peak at 2900~\AA\ also appears
to be common to these objects, along with a bluer peak near 3150~\AA.
Shortward of $\sim$2800~\AA, the SNe Ia diverge from the core-collapse
objects.  This dropoff in flux, along with a pair of deep absorption
features near 2450~\AA\ and 2600~\AA, is caused by iron-group elements,
primarily \ion{Co}{2} \citep{kirshner93,sauer08}.  The lack of these
features in  \afx\ is evidence that the abundance of newly-synthesized
$^{56}$Ni (which decays to $^{56}$Co) near the photosphere is much
lower than in a SN~Ia.

 One
purpose of these comparisons is to demonstrate the apparent broad
similarity of the NUV spectra of \afx\ to those of SNe coming 
from three very different progenitors: a bare core of a massive star
(SN 1994I: \citealt{wheeler94,iwamoto94i}), a partially-stripped
massive star in a binary system that retained part of its hydrogen
envelope (SN 1993J: \citealt{nomoto93j,pods93j,woosley93j}), and the
thermonuclear explosions of white dwarfs (SNe 2011fe and 2011iv:
\citealt{nugent11fe,foley11iv}). SNe 2011fe and 2011iv have some
spectral differences from each other near 3000~\AA, exhibiting the 
potential for spectroscopic variation in objects with similar
progenitors.  Other than the lack of cobalt near the photosphere, the
remaining differences between \afx\ and these other objects are
relatively minor.

In addition, the overall UV spectral slope of \afx\ is very similar to
those of SNe 1993J and 1994I, after correction of both by
$E(B-V)=0.30$~mag of reddening \citep{richmond94,sauer06}.  The fact
that the only SNe in Figures \ref{uvcomp} and \ref{uvcomplog} that
have similar spectral features to \afx\ also have similar SED shapes
provides further evidence that the extinction of \afx\ is not large.
If the \afx\ spectra were corrected for a large amount of reddening,
they would be bluer and we would expect different spectral features
to be present at the higher implied temperatures.

We further explore the spectra using the SN spectrum synthesis code
\texttt{SYN++} with 
\texttt{SYNAPPS} \citep{thomas11} to fit the near-maximum-light
GMOS/FIRE spectra.  These codes are based on the same 
assumptions as the original \synow\ (e.g., \citealt{branch02})
treatment of resonant-scattering lines in the Sobolev approximation
above a sharp photosphere that emits as a blackbody (BB).  The atomic
and ionic species are assumed to have levels populated in local
thermodynamic equilibrium at some excitation temperature.  \synow\ is
a tool designed for line identifications and is not a full
self-consistent spectral model.

\begin{figure}
\epsscale{1.2}
\plotone{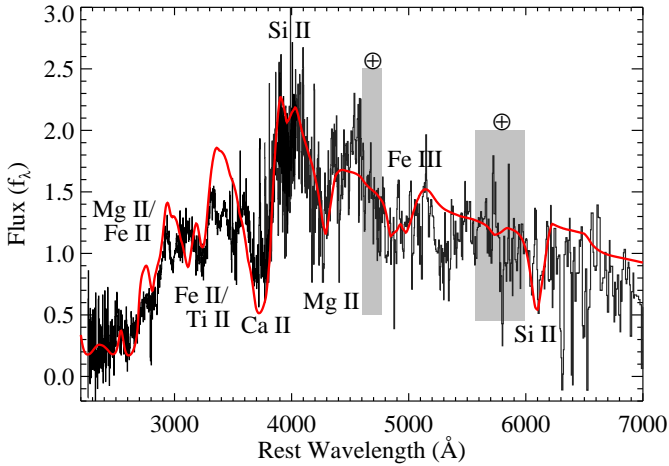}
\caption{\synow\ fit for \afx\ (red) compared to the $-$2.8/$+$0.1~d
  GMOS/FIRE combined spectrum.  Ions dominating major spectral
  features in the \synow\ model are indicated.  The gray bars labeled
  with a $\earth$ symbol represent the regions of strong telluric
  absorption. 
}
\label{synplot}
\end{figure}

Our best \synow\ model is shown in Figure~\ref{synplot} and
contains seven ions.  We assumed equal BB continuum and excitation
temperatures of $10,200$~K and a photospheric velocity of 
$11,000$~\kms.  The deep absorption trough near 3700~\AA\ is dominated
by \ion{Ca}{2} H\&K.  \ion{Fe}{2} and \ion{Ti}{2} combine to produce
several of the wiggles observed in the range $3000-3500$~\AA.
\ion{Mg}{2} naturally explains the 
observed absorption near 4250~\AA, while also producing absorption
shortward of $\sim$2700~\AA.   \ion{Si}{2} contributes to both the
blue wing of the deep \ion{Ca}{2} feature and the notch near 3950~\AA.
The feature near 3950~\AA\ is present in the overlap between the GMOS
and FIRE data and has a consistent strength that is significantly
larger than in the synthetic spectrum.  With \synow, we have to be
careful not to overproduce absorption by the \ion{Si}{2} $\lambda$6355
doublet because no strong feature is seen in the data, although the
S/N and the spectral resolution become very poor in the observed-frame
$H$.  We introduce \ion{Fe}{3} to help reproduce the wiggle near
5000~\AA, but the identification is not unique and its presence is not
required.  In addition, we include \ion{Cr}{2} to help suppress the UV
flux.  That ion does not contribute sufficiently strongly to 
any specific feature for us to make a positive identification,
but it has been included in previous \synow\ models of SNe
Ic \citep{millard99} because of its expected strength in
either helium or carbon/oxygen-rich SN ejecta at these temperatures
\citep{hatano99}.

We also consider several other species that were not included in the
final fit. As discussed above, \ion{Co}{2} produces strong NUV
absorptions in SNe Ia that are not present in \afx.  \ion{Na}{1} D is
commonly seen in all types of SNe, but at this redshift it would fall
in the strong telluric absorption between $J$ and $H$, so we have no
constraint on its presence from the observations.  We also tested
\ion{Sc}{2} because of its use in prior \synow\ analyses of SNe Ib
\citep{branch02}.  Scandium results in an additional absorption
minimum near $\sim$3500~\AA\ similar to one seen in our data, but does
not produce a sufficiently strong improvement in the model to justify
its inclusion, particularly given the expectation that it should only
become important at lower ejecta temperatures \citep{hatano99}.
There is also an apparent absorption feature near 5300~\AA\ 
that is not fit by the model.  \ion{S}{2} can produce a feature near 
that wavelength, but the identification is not certain.
   Lines of hydrogen or helium would be very important if
present, but there is no evidence for either, although the low S/N of
our NIR data precludes strong statements about helium.  The SN
2005ap-like SLSNe have lines of singly-ionized carbon and
oxygen in their spectra at early times (e.g., \citealt{quimby11}),
with possible contributions from doubly-ionized CNO elements
\citep{10gx}, but those lines do not appear to be present in our
\afx\ spectra. 

Overall, we can reproduce the shape of the spectrum in the rest-frame
optical reasonably well, but the NUV is more problematic.  Our
synthetic spectrum has features at many of the same wavelengths as the
real data, but the overall shape in the NUV is less well fit.  Adding
additional ions to the fit might potentially help 
the UV shape of the synthetic spectrum, but would not be well
motivated or add any additional insight.  Alternatively, the radiative
transfer in the UV is known to be complex and we could be encountering
the limitations of \synow.  The most
important point is that the ions we have
used are typical for generic fits to SNe Ia and Ic spectra (e.g.,
\citealt{branch02}), but differ from those identified in most
SLSNe-I spectra near maximum light \citep{quimby11,laura,06oz,bzj}.

\section{Light Curve and Energetics}

\begin{figure}
\epsscale{1.1}
\plotone{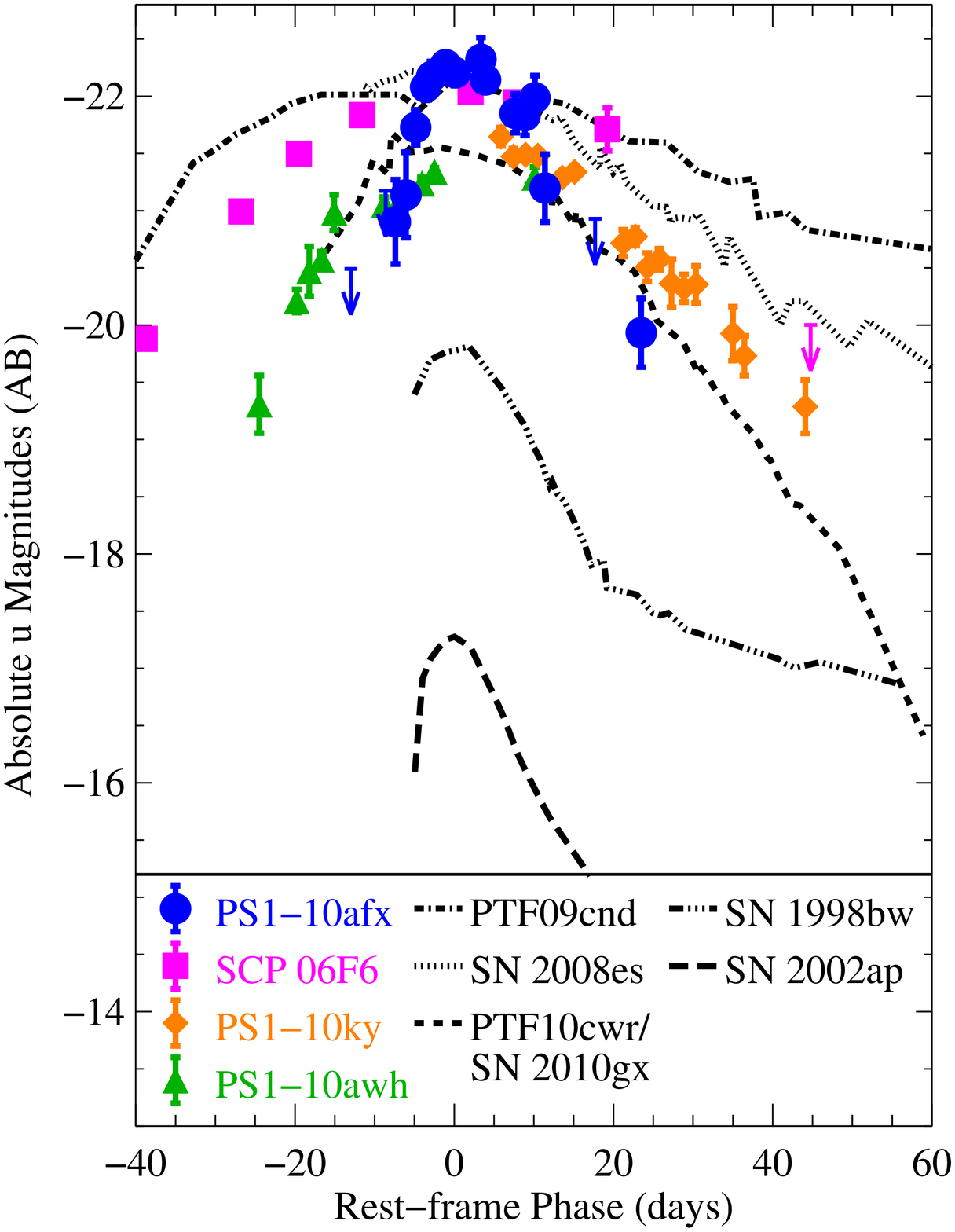}
\caption{Absolute $u$ light curves of \afx\ and other SNe.
  The 
  observed bands and wavelengths in the rest frame are: \afx\ (\zps,
  3525~\AA), SCP 06F6 (F850LP, 3883~\AA; \citealt{barbary}), PS1-10ky
  (\ips, 3848~\AA; \citealt{laura}), PS1-10awh (\ips, 3944~\AA;
  \citealt{laura}), PTF 09cnd (rest-frame $u$ from
  \citealt{quimby11}), SN 2008es ($B$/$g$, 3604~\AA;
  \citealt{miller08es,gezari08es}), SN~2010gx/PTF~10cwr ($g$,
  3770~\AA; \citealt{10gx,quimby11}), SN~1998bw ($U$, 3663~\AA;
  \citealt{galama}), and SN~2002ap ($U$, 3663~\AA;
  \citealt{foley02ap}).
}
\label{uplot}
\end{figure}

Existing light curves of SLSNe are heterogeneous, and the objects have
been found at a wide range of redshifts, making direct comparisons
difficult.  Following previous work \citep{quimby11,laura,bzj}, in
Figure~\ref{uplot} we compare the rest-frame $u$ light curve of
\afx\ (corresponding to observed-frame \zps) to several of the most
luminous SN 2005ap-like
SNe in the literature.  In addition, we include the hydrogen-rich
SN~2008es because its light curve is similar to the others and it did
not develop strong Balmer lines until well after maximum light
\citep{miller08es,gezari08es}.  For comparison, we also plot the $U$
light curves of the broad-lined SNe Ic 1998bw and 2002ap, two objects
of normal luminosities \citep{galama,foley02ap}.  Given the uncertain
SEDs of many of these objects, we do not perform a full
$k$-correction, but instead correct the observed AB magnitudes in
filters whose effective wavelengths are close to $u$ in the observed
frame by $m-M=5\log($d$_L$($z$)/10 pc) $-$ 2.5 $\log$(1+$z$), where
d$_L$ is the luminosity distance.  To be conservative, we assume that
\afx\ is unaffected by  
reddening here and in all following discussion, despite the strong
\ion{Mg}{2} absorption along the line of sight and the inference
that the host galaxy SED requires some degree of internal extinction.
We presented some evidence above that the reddening to the SN is not
large, but it is possible that we have underestimated the peak
luminosity of this object.

It is apparent from Figure~\ref{uplot} that at peak \afx\ was slightly
more luminous in $u$ ($M_{u} \approx -22.3$~mag) than any previous
SLSNe.  However, by $\sim$10~d on either side of maximum light,
\afx\ falls below all the others.  This emphasizes the high peak
luminosity and fast evolution of the light curve.  These objects have
rather different SEDs, so comparisons of the energetics require
constructing a bolometric light curve.

\begin{figure}
\epsscale{1.2}
\plotone{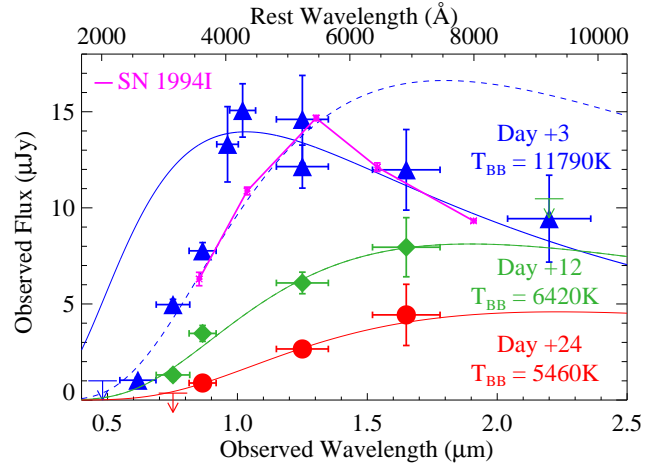}
\caption{SED of \afx\ at three epochs, phases of $+3$ (blue triangles),
  $+$12 (green diamonds), and $+$24~d (red circles).  The solid lines
  are BB fits to the SED at each epoch.  At a phase of $+$3~d, the
  blue dashed 
  line is a fit to all data points, while the solid line excludes
  \gps\rps\ips\zps\ from the fit.
Reported BB temperatures are in the rest frame.
The magenta line shows the SED of SN 1994I near maximum light shifted
to the redshift of \afx\ and arbitrarily scaled to match the flux
level.  The 1994I data points are the $UBVRI$ photometry on 1994 April
5.39 from \citet{richmond96} corrected for $E(B-V)=0.3$~mag
\citep{sauer06} of reddening.
}
\label{sedplot}
\end{figure}

To proceed further, we first construct SEDs at three epochs near the
times of our NIR observations, at phases 
of $+3$, $+12$, and $+24$~d.  We fit third-order polynomials to the
observed photometry in each \rps\ips\zps\ filter around each SED epoch
to estimate the flux.  We then estimate the error bars on these
fitted fluxes by generating Monte Carlo realizations of the data where
we repeatedly randomly adjust the observed fluxes by drawing from a
normal distribution having a width of the measured errors, refitting,
and determining the variance of the interpolated flux at the 
desired epoch.  We perform a similar procedure whenever
we need to interpolate fluxes to a common epoch.  For the SEDs at
phases of $+3$ and $+24$~d, we use the actual measured \yps\ and NIR
fluxes.  For the $+12$~d SED, the observed NIR fluxes were quite
noisy, so we again interpolate the fluxes in $J$ and $H$ by fitting
second-order polynomials to the observed light curves to generate the
points on the SED.  The resulting SED evolution is shown in
Figure~\ref{sedplot}.

We then fit BB spectra to the SEDs at each epoch.  Our $+3$~d SED has
the most complete information.  The dashed line in 
Figure~\ref{sedplot} shows the result of a BB fit to the full SED.
The derived BB temperature is $\tbb\approx6800$~K, but clearly the
curve is a poor fit to the data at rest wavelengths longward of
$\sim$5000~\AA.  If instead we restrict the fit to \yps\ and the NIR
bands, the 
BB temperature is not well constrained ($\tbb=12,000\pm2,000$~K) 
due to a lack of points blueward of the peak.  The formal best fit is
plotted as the solid blue line in Figure~\ref{sedplot} and provides a
much better fit to the NIR data points (by construction), but
significantly overestimates the rest-frame UV flux.  This is to be
expected if line blanketing in the NUV provided by iron-peak elements
supresses the UV flux, as is typical for SNe Ic.  For
example, the shape of the SED of the normal SN Ic 1994I near maximum
light (Figure~\ref{sedplot}) is similar to
that of \afx, but the detailed model fits by \citet{sauer06} find an
underlying BB temperature of 9230~K at the same epoch.  

\begin{figure}
\epsscale{1.2}
\plotone{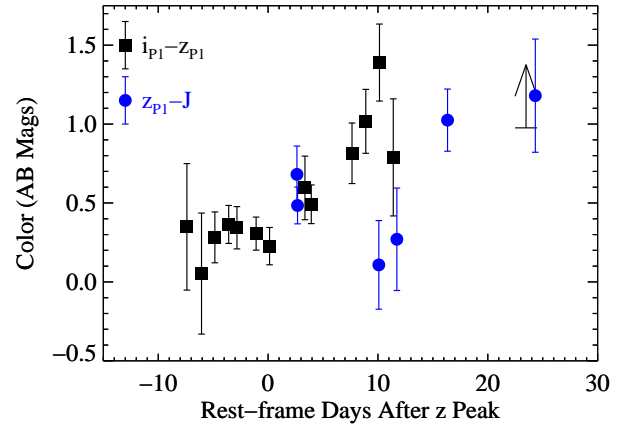}
\caption{Observed color evolution of \afx.  At this redshift, \zps\ is
  near 
  rest-frame $u$ band and $J$ is near $V$, while \ips\ is in the
  NUV. Note the flat $\ips - \zps$ color prior to maximum light.
}
\label{colorplot}
\end{figure}

The data at phases $+12$ and $+24$~d are less complete, so we fit a
simple BB to all of the points on the SED and do not correct for line
blanketing, which would only be expected to increase as the ejecta
cool.  The derived BBs show the ejecta clearly cooling from
$\sim$$12,000$~K at a phase of $+3$~d to $\sim$$5500$~K at $+24$~d,
although \tbb\ should only be regarded as lower limits to the
effective temperature ($T_{\mathrm{eff}}$) on the later two dates.  Independent
of the details of  line blanketing, we can directly see the SED
becoming redder, and hence implying cooler temperatures, by
looking at the color evolution.  In Figure~\ref{colorplot}, we
construct \ips$-$\zps\ and \zps$-J$ color curves.  In each case, we
interpolate the light curve in the bluer band to the epochs of
observation for the noisier redder band.  The \ips$-$\zps\ color is 
consistent with a constant (0.30$\pm$0.05~mag) before
maximum light, but both colors became redder with time after
maximum. 

We use the BB fits to construct a bolometric light curve, starting
with the near-maximum-light SED (phase $+$3~d) shown in 
Figure~\ref{sedplot}.  A trapezoidal integration of the observed
fluxes from \rps\ to $K$ gives a minimum bolometric luminosity
(\lbol) near peak of (3.6$\pm$0.2)$\times10^{44}$~erg~s$^{-1}$.
To account for flux emitted redward of our observations, we add
the tail of the BB
and extend the trapezoidal integration to long wavelengths, which
provides only a 14\%\ upward correction to the total flux.  We
repeat the process for the other two SED epochs.  After including
the BB tails, the derived bolometric luminosities are (4.1$\pm$0.2),
(2.2$\pm$0.2), and (1.1$\pm$0.3)$\times10^{44}$~erg~s$^{-1}$ at phases
of $+$3, $+$12, and $+$24~d, respectively.   

\begin{figure}
\epsscale{1.2}
\plotone{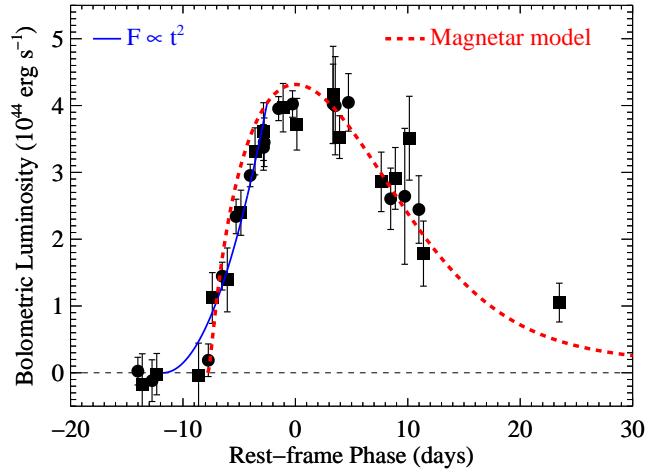}
\caption{Bolometric light curve of \afx.  The observed \ips\ (circles)
  and \zps\ (squares) light curve points were multiplied by a
  time-dependent bolometric correction factor (see text for details).
  The plotted error bars only represent the nominal photometric errors
  and do not include any contribution from the systematics of
  converting to bolometric luminosities. The solid blue line shows a
  fit to the early-time flux that rises as $t^2$.  The red dashed
  line shows a model for a SN powered by 
  the dipole spindown energy from a magnetar, following
  \citet{kb10}.  We assume an opacity of 0.2~cm$^2$~g$^{-1}$, 1~\msun\
  of ejecta, 10$^{51}$~erg of kinetic energy in the initial SN, an
  initial magnetic field strength of 6$\times$10$^{14}$~G, and an
  initial magnetar spin period of 2~ms.  This model predicts high
  temperatures and velocities that are in contradiction of the data
  despite the apparent good fit.  See Section 6.2 for further
  discussion.
}
\label{boloplot}
\end{figure}

Although we lack NIR data at other epochs, we can use these fits to
estimate the bolometric light curve in combination with the
well-sampled PS1 light curves.  For each of our three SED epochs, we
define a 
multiplicative bolometric correction factor to obtain \lbol\ from
$\nu$$L_{\nu}$ in the \zps\ band.  Motivated by the lack of strong
color evolution before maximum light, we apply the bolometric
correction factor from $+$3~d to all pre-maximum \zps\ data points.
For the post-maximum data, we use a smoothly varying time-dependent
correction factor to evolve from the $+$3~d value to the $+$24~d
value.  We repeat the process for the \ips\ band, except that we
use a smaller (but constant) bolometric correction factor before
maximum light than the derived $+$3~d value to better match
the \zps\ results and account for the small amount of
\ips$-$\zps\ color evolution between $+$3~d and the pre-maximum
data.  The combined bolometric light curve is 
plotted in Figure~\ref{boloplot}.  A third-order polynomial fit to the
peak of the bolometric luminosity curve gives a maximum value of
(4.1$\pm$0.1)$\times 10^{44}$~erg~s$^{-1}$, equivalent to
1.1$\times$10$^{11}$~L$_{\sun}$, not including the uncertainty in the
bolometric correction factors (or potential extinction). 

To check that our derived maximum bolometric luminosity of \afx\
is realistic and that our procedure for extrapolating the SED
to the NIR has not introduced any significant error, we can take
advantage of the fact that the shapes of the spectra we have of
\afx\ and the derived SEDs are very similar to those of a SN Ia near
maximum light (cf. Figure~\ref{uvcomplog}).  One of the most
distant spectroscopically confirmed SNe Ia known, HST04Sas, is at
$z=1.39$ \citep{riess07}, consistent with the redshift of \afx. 
Interpolating the F850LP ($\sim$$z$ band) light curve of HST04Sas to a
phase of zero days produces an estimate of the observed magnitude
of $m_{F850LP}(t=0) = 25.5$~mag (AB), while fits to \afx\ show that
\zps$(t=0) = 21.7$~mag.  Assuming that the similarity of the SEDs
requires no further correction factors and a typical SN Ia peak
bolometric luminosity of $\sim$1.2$\times$10$^{43}$~erg~s$^{-1}$
(e.g., \citealt{max06}), the 3.8~mag difference between the objects
corresponds to a peak bolometric luminosity for \afx\ of
$\lbol\approx4\times$10$^{44}$~erg~s$^{-1}$, in excellent agreement
with our value derived above.

In fact, the peak of $\mbol =-22.8$~mag is more luminous than any of
the hydrogen-poor objects collected by
\citet{galyam12}\footnote{For consistency, we note that using the same
  cosmology as Gal-Yam would result in $\mbol =-22.9$~mag.}.  This
statement is somewhat uncertain due to the heterogeneous and
fragmentary nature of the available data for SEDs and bolometric
correction factors for the objects in the literature.  SNe 2008es
\citep{miller08es,gezari08es}, 2005ap \citep{quimby05ap}, and SCP06F6
\citep{barbary,quimby11} all have 
similar peak bolometric luminosities to \afx, but the exact comparison
depends on the treatment of the bolometric correction for these
UV-luminous sources.  The only SLSN of any type with a clearly higher
peak luminosity is the peculiar hydrogen-rich transient CSS100217
\citep{drakesn}.  However, the SN 
nature of that object is still questionable due to possible confusion
with the active galactic nucleus of its host galaxy.

In addition to the high peak luminosity of \afx, another striking
fact about the light curve is the rapid time evolution.
Despite the rolling survey nature of PS1, our first detection is at a
phase of only $-7.4$~d, with non-detections prior to that
(Figure~\ref{boloplot}).  It is traditional in SN studies to find the
rise time by fitting a fireball model of the form
$\lbol\propto t^{2}$ for $t>t_0$ at early times, where $t_0$ is the
explosion time. For \afx, this gives a $t_0 = -11.8\pm0.7$~d
(Figure~\ref{boloplot}).  Allowing the power-law index to vary gives a
best fit with $\lbol\propto t^{0.9\pm0.5}$ and $t_0=-8.7\pm1.9$~d.
Regardless of the exact parameterization, the rise in \zps\ is a
factor of 3.5 in flux in the 7.4~d before peak. This apparent rise
time is much faster than the fastest known SLSNe
(Figure~\ref{uplot}; \citealt{quimby11,10gx,06oz,bzj}). 

The decay timescale is also unusually fast.  The FWHM of the
bolometric light curve is only 18~d.  The usual  
parameterization of SN light curve shapes, \dm, is the decline in
magnitudes in 15~d after maximum light.  The bolometric light curve of
\afx\ has \dm=0.95 mag.  By comparison, only SN~1994I has a
faster \dm($R$) in the collection of SNe Ib/c light curves of
\citet{drout11}, although several additional objects had \dm($V$) that
are comparable.  Figure~\ref{uplot} demonstrates that in $u$,
\afx\ also has a faster light curve decay than SLSNe of similar peak
luminosity. 

Integrating the observed bolometric light curve from the first
detection at a phase of $-$7.4~d to the last detection at $+$24~d
provides a lower limit on the emitted energy of
$\sim$7$\times$10$^{50}$~erg (over only 31~d).  Reasonable
extrapolations of the light curve to later times will increase
the total to $\sim$10$^{51}$~erg, comparable to most other
hydrogen-deficient SLSNe, but not notably high (e.g.,
\citealt{quimby11,laura}).  The fast timescale of \afx\ compensates
for its high peak luminosity.

\section{Comparison to Models}

The difficult constraints placed by the rise time on any model can be
seen from the inferred emitting radius.  Assuming $T_{\mathrm{eff}}
\approx 11,790$~K as derived above, the BB radius (\rbb) of the
photosphere  at maximum light is 5.5$\times10^{15}$~cm.  This is
larger than for other hydrogen-poor SLSNe, which typically have
inferred radii of $\sim$10$^{15}$~cm \citep{quimby11,laura}, as 
expected from the higher luminosity and lower \tbb\ of \afx.  If
$T_{\mathrm{eff}}$ is actually as low as   
the color temperature of $\sim$$6800$~K, then the inferred radius
at maximum light increases to 1.6$\times10^{16}$~cm.  

The most conservative set of numbers at maximum light
(\rbb =5.5$\times10^{15}$~cm, $t_0$=$-$11.8~d) requires an
expansion velocity of $\sim$50,000~\kms\ if $t_0$ is
the true explosion time, in obvious conflict with the observed 
photospheric velocities seen in the spectrum.  Material moving at the
\synow-derived velocity of $11,000$~\kms\ will take 58~d to reach
that radius, implying an explosion $\sim$50~d before our first
detection.  There is an observed example of such an effect with the
SLSN~2006oz.  It exhibited a precursor plateau with a luminosity of
$\sim$2$\times$10$^{43}$~erg~s$^{-1}$ prior to rising to the main peak
of the light curve, which \citet{06oz} interpreted as being due to a
recombination wave in an oxygen-rich CSM.  A weighted average of the
\afx\ non-detections between phases $-15$ and $-8$~d produces a mean
luminosity of (0.2$\pm$1.2)$\times$10$^{43}$~erg~s$^{-1}$.  At the
2$\sigma$ level, this non-detection is consistent with a hypothetical
plateau being present at the level seen in SN~2006oz, but unobserved
due to the limiting magnitudes of our survey. 

At phases $+$12 and $+$24~d, the
\lbol\ quoted above and \tbb\ shown in Figure~\ref{sedplot}
combine to produce BB radii of 1.4 and 1.3$\times10^{16}$~cm,
respectively, although these are likely overestimated due to the
underestimate of $T_{\mathrm{eff}}$ caused by line blanketing.  Notably, these
values are consistent with the photosphere begining to recede
through the ejecta (in both the physical and comoving
senses) as they become optically thin.  By contrast, the BB
radii inferred for the interacting SLSNe 2006tf and 2006gy
increased linearly with time at the same rate as the velocities
measured from the spectra, consistent with an expanding
optically-thick shell \citep{smith06tf,smith06gyb}.  Although the
measurement errors make it less certain, this also appears to be true
for at least some SN 2005ap-like SNe (e.g., \citealt{laura,bzj}).

\subsection{Models Powered by Interaction}

The kinetic energy of a powerful SN explosion provides an attractive
reservoir of energy to power SLSN light curves, but only if that
energy can be 
efficiently converted to radiation.  For \afx, this already requires
an initial SN energy of at least a $few\times$10$^{51}$~erg in order
for the object to radiate 10$^{51}$~erg and still have ejecta
velocities as high as those observed.  This generally requires a CSM
with a mass comparable to that of the ejecta and a radius
comparable to the photospheric radius at maximum light in
order to maximize efficiency and not lose energy to adiabatic
expansion (e.g., \citealt{gb12}).  Another attraction of interaction
models is that the SN can explode well before the start of the optical
rise of the SN if the CSM is sufficiently dense to trap radiation from
the shock \citep{ci11}.  This could greatly alleviate the conflict
described above between the rise time, ejecta velocity, and
photospheric radius.

The required CSM densities in these models are much larger than for
well-studied examples 
at lower luminosity.  \citet{ci11} have identified two regimes of
interest when the CSM is this dense.  If the radiation from the shock
can escape while the shock is still interacting with the CSM, then the
light curve will be broad and exhibit a slow decay after maximum
light.  Their model works well for the light curves of SLSNe like
2006gy.  An additional consequence of this model is that there should
be spectral signatures of the ongoing CSM interaction, such as
emission lines from the unshocked CSM (e.g., \citealt{chugai01}).
Other luminous SNe IIn in the literature, such as SNe 2008fz
\citep{drake08fz} and 2008am \citep{08am}, with their blue continua
and strong Balmer emission lines, can also be fit by similar models. 

The much faster light curve of \afx\ and lack of strong emission lines
imply a different regime, one where the outer radius of the CSM is
comparable to or smaller than the diffusion length of photons from the
shock.  In this framework, the optical light we detect is then
associated with the 
breakout of the shock from the outer radius of the CSM.  In the
context of a dense wind with a density profile that scales with radius
as $\rho = D r^{-2}$ out to a maximum radius of $R_w$, where $D$ is a
density normalization parameter, the total mass of the wind is $M=4 \pi
R_{w} D$.  In the models of \citet{ci11}, $D$ can be estimated
from the diffusion timescale as $D = 1.28 \times 10^{16}$ ($t_d$/days)
g~cm$^{-1}$, if we assume an ionized He-rich composition.  We can
approximate the diffusion timescale $t_d$ by the 
rise time and set $R_w$ to the BB radius at maximum light to
produce an estimate of the wind mass of $5-10$~\msun.  These values
are only approximate, as \citet{gb12} have shown that in the regime of
interest, the analytic approximations are outside the range of
validity and require true hydrodynamic calculations.

Models with shock breakout occurring in a dense, truncated wind can
produce 
bolometric light curves with the desired peak luminosities and time
scales, but it is unclear if this can be done consistently with the
other information about the object.  \citet{gb12} were able to
approximately match the light curve shapes of the SLSNe 2010gx and
2005ap, but the derived \tbb\ were hard to match to the
observations.  A theoretical caveat is that the observed
color temperature could deviate from $T_{\mathrm{eff}}$ if the shock is not in
equilibrium when it breaks out \citep{nakar}.  This deviation is
caused by the nature of the opacity, and usually implies that
the color temperature is hotter than the true $T_{\mathrm{eff}}$ (e.g., see
\citealt{moriya06gy} for models of 2006gy), which would in turn
require that the photosphere be at even larger radii than we have
found. This effect might then conflict with the fast
timescale of the light curve, which otherwise is indicative of a
more compact CSM \citep{ci11,gb12}, but detailed modeling with
radiative transfer is needed to correctly evaluate these concerns.

Further, a key component of the rising SN luminosity in
the shock breakout models of both \citet{ci11} and \citet{gb12} is
that the temperature rises to a 
maximum near the peak of the light curve and declines thereafter.
Although we lack full SED information on the rise of the light curve
for \afx, the observed $\ips - \zps$ color shows no evolution prior to
the peak of the light curve (Figure~\ref{colorplot}).  Despite the
interpretation uncertainties introduced by the possibilities of line
blanketing or deviations from equilibrium, the evolution of \tbb\
after maximum light results in a noticeable reddening of the $\ips -
\zps$ color.  If \tbb\ were rising by a similar amount while the
light curve rose to its peak, we would have been able to detect it.

A separate uncertainty is the origin of this CSM, which could not be
produced by steady stellar winds due to the high required mass-loss
rates and the unexplained truncation at $R_w$.  Most theoretical work
has focused on winds, but the material could also be in a shell
\citep{smithmccray,mt12}.  \citet{woosley07} invoked a pulsational
pair instability to produce shells of matter at the appropriate
distances, but these require very massive stars.  Other possibilities
include common envelope ejection \citep{chevalier12} and instabilities
in the late stages of stellar evolution \citep{qs12}.

Another challenge for CSM interaction being the dominant contributor
to the luminosity of \afx\ is the appearance of the spectra.
Interaction-powered SNe convert the kinetic energy of the explosion
into UV/optical continuum radiation, frequently reprocessing some of
the continuum into strong emission lines from the dominant
constituents of the CSM external to the expanding shock.  The
spectral comparisons in Figure~\ref{uvcomp} show several examples of
objects consistent with this picture.
However, \afx\ is even more luminous than those objects and yet shows
P-Cygni features originating in the SN ejecta, including a deep
\ion{Ca}{2} absorption comparable to that seen in SN 1994I.  The lack
of strong emission lines from the CSM can be explained if the material
has a sharp outer radius, but interaction 
sufficiently luminous to dominate the light curve should dilute and
weaken spectral features from the ejecta due to the ``top-lighting''
effect \citep{branchtop}, if not completely hide emission from the
unshocked ejecta. For example, PTF~09uj is a candidate for being a
lower-luminosity version of a shock breakout through a dense
(hydrogen-rich) wind and, as expected, its spectra are very blue and
lack strong P-Cygni features \citep{ofek09uj}.  It is not clear how
strong CSM interaction could dominate the luminosity of \afx\ without
leaving spectral signatures. 

In summary, existing CSM interaction models can match the gross
photometric properties of \afx\ (peak luminosity, timescales), but
only if the models have the freedom to assume the necessary CSM
structure.  The origin of such unusual CSM is not explained {\em a
  priori} in these models.  In addition, the details of the SED,
color evolution, and emitted spectra appear to be in conflict with the
observations, although more detailed radiative transfer calculations
are necessary to be confident in the model predictions, which are
largely based on BB assumptions.

\subsection{Internal Energy Sources}

Normal hydrogen-deficient SNe have light curves powered by the
diffusion of energy deposited by the radioactive decay of $^{56}$Ni
and $^{56}$Co.  The light curves of most SLSNe-I appear to fade too
rapidly relative to their peak luminosities to have
radioactively-powered ejecta \citep{quimby11,10gx,laura}, and \afx\ is
no exception.  With a rise time of 12~d, the peak bolometric
luminosity requires a nickel mass of $\sim$14~\msun.  However, making 
standard assumptions about the structure of the ejecta
\citep{arnett82,03jd,drout11} allows us to use the photospheric
velocity of 11,000~\kms\ and light curve timescale of 12~d (assuming
that the diffusion timescale is comparable to the rise time) to
estimate the ejecta mass as $\sim$2~\msun, clearly in contradiction of
the high nickel mass.  

Another potential power source suggested for SLSNe is the spindown
energy from a newly-born magnetar
\citep{maeda05bf,woosley10,kb10,dessart_mag}.  The high 
luminosity and fast rise of \afx\ stretch the existing models
to the bounds of plausibility.  Inspection of Figures~4 and 5 of
\citet{kb10} shows that the light curve of
\afx\ requires a magnetar solution to have a very fast initial spin
period ($p\approx1-2$~ms, near the breakup speed of a neutron star), a
low ejecta mass ($\sim$1~\msun), and a high magnetic field
($\gtrsim few\times$10$^{14}$~G).  This is because the high
luminosity requires a high spindown power, while the fast light curve
requires both fast spindown times and short diffusion timescales.  We
examined magnetar-powered light 
curves using the formalism of \citet{kb10} and found parameters that
produced light curves that approximately matched the shape of the
bolometric light curve of \afx.  One such model is plotted in
Figure~\ref{boloplot}.

However, it is not clear whether this fit should be regarded as
more than a numerical curiosity because the physical assumptions
underlying the model break down.  In the parameter range
appropriate for \afx, the magnetar dipole spindown energy
(5$\times$10$^{51}$ ($p$/2~ms)$^{-2}$ erg) dominates over the initial
SN kinetic energy (assumed to be 10$^{51}$~erg).
The \citet{kb10} models make an ansatz that the entire spindown energy
of the magnetar is simply thermalized spherically at the base of the
SN ejecta and deposited into the internal energy.
Because this process accelerates and heats the ejecta, the final
kinetic energy, the peak luminosity, and the temperature are coupled.
The model plotted in Figure~\ref{boloplot} predicts expansion
velocities of $\gtrsim$25,000~\kms, in contradiction of the observed
spectrum.  In addition, the model predicts a radius at maximum light
(from the ejecta velocity and rise time) that when combined with the
luminosity implies a BB temperature in excess of $20,000$~K (with
strong evolution at other times).  This contradicts the 
observed \tbb\ near maximum light and the constant red
\ips$-$\zps\ color during the rise of the light curve.  We find that
increasing the assumed SN kinetic energy relative to the spindown
energy does not alleviate these issues because of the difficulty in
simultaneously satisfying the constraints from the 
expansion velocity seen at maximum light (which implies a normal
energy-to-mass ratio, and hence high mass if the initial SN energy is
high) with the fast diffusion timescale (which requires a low mass).

The models of \citet{woosley10} start from massive star progenitors
and proceed through the explosions, but they only probe slower
rotating, and hence less energetic, magnetars than those required for
\afx. \citet{dessart_mag} also examined explosion models for Type Ib/c
SNe with the addition of a central energy source that mimicked the
effect of a magnetar.  They followed the effects of a central energy
source on the ejecta produced by the explosion models and generated
light curves and spectra.  If the central energy source deposited
$\sim$10$^{51}$~erg of energy, the resulting SNe had high velocities
and were blue at maximum light, which appears promising for explaining
the SN 2005ap-like class of SLSNe, but is very different from \afx.
As noted above, the light curve of \afx\ requires even more
energy input from the magnetar. Therefore, we disfavor magnetar models
at this time. 

\citet{dexter} have proposed a scenario that shares some similarities
with the magnetar model in that their model has a post-explosion
internal energy source, in this case provided by fallback accretion
onto a newly formed compact remnant, likely a black hole.  Their suite
of models can 
only match the fast timescale and high peak luminosity of \afx\ if the
progenitor is a blue supergiant and the initial SN explosion energy is
extremely low with an ejecta mass of $\sim$1~\msun.  Their
hydrogen-poor models do not reach the same peak luminosities, which
is a problem for \afx.  In addition, these models have the same
problem as the magnetar ones in simultaneously satisfying the high
peak luminosity and relatively cool photospheric temperature.

\subsection{Asphericity}

In all of the preceding discussion, we have necessarily assumed
spherical symmetry in order to convert our observations into intrinsic
quantities.  However, core-collapse SNe are all known to be aspherical
at some level (e.g., \citealt{ww08}).  The most basic effect of
asphericity is to increase the uncertainty in \lbol\ and dilute the
connection between observables such as the rise time and physical
quantities such as the diffusion timescale.  

The broad-lined SNe Ic associated with GRBs have been extensively
studied in the context of aspherical models, with different studies
coming to conflicting conclusions for the same object (e.g., SN
1998bw: \citealt{hoflich99,maeda06}).  The models of \citet{maeda06}
for SN 1998bw predict that lines of sight along the major axis of a
bipolar outflow will see light curves with faster rise times and
higher luminosities than average, perhaps indicating a general trend
with relevance for \afx.  However, despite the high explosion energies
($>$10$^{52}$~erg) and fairly high nickel masses ($\sim$0.4 \msun)
assumed in their models, the maximum peak luminosity was still
an order of magnitude lower than for \afx.  In addition, the spectral
features of \afx\ are not as blended as those of broad-lined SNe Ic
(Figure~\ref{uvcomp}).  We conclude that an aspherical SN 1998bw-like
model is unlikely to produce an object that resembles \afx\ unless
some other ingredient is added to the models.

Another possibility in a CSM interaction scenario is that the material
with which the SN is interacting is distributed in an aspherical
fashion, perhaps allowing a distant
observer to see both the luminous continuum emission from the
interaction region and the SN ejecta.  However, such models still
require the combination of the covering fraction of the CSM, the
radiative efficiency factor, and the SN kinetic energy to be
sufficiently high to produce the $\sim$10$^{51}$~erg of optical
emission that we measure.  The SLSN models of
\citet{metzger10} invoke remnant protostellar disks around massive
stars, but do not simultaneously fit the high peak luminosity and the
fast timescale of \afx.  In addition, such a model has similar
difficulties as spherical CSM interaction models in explaining the
strength of the P-Cygni features in the observed spectra if there is
continuum emission from an interaction region that dilutes the
strength of the spectral features from the ejecta. 

\subsection{Is \afx\ a Lensed SN?}

We now consider whether gravitational lensing could result in an
object with the spectrum and light curve shape of a normal SN, but
artificially boosted in luminosity.  However, the peak luminosity of
\afx\ is a factor of $\gtrsim$50 higher than normal SNe Ic, requiring
an extremely high magnification factor despite the lack of an
obvious lens.  While this work was being refereed, a preprint by 
\citet{quimbylens} expanded on this hypothesis further and 
proposed that \afx\ was instead a normal SN Ia and therefore had 
to be lensed by a magnification factor of $\sim$20.

There are several constraints against the lensing hypothesis, starting with
the lack of an apparent lens.  In standard $\Lambda$CDM cosmologies, 
the optical depth to lensing with amplification factors $>$10 is very small, 
10$^{-5}$ $-$ 10$^{-6}$ for sources at the redshift of \afx\ 
(e.g., \citealt{hilbert08,taka11}), and the mass distribution of the lenses
for those extreme magnifications is peaked at cluster-scale haloes of 
$\sim$10$^{14}$~\msun\ \citep{hilbert08}.
However, there is no foreground galaxy cluster visible in the
PS1 images, as \citet{quimbylens} also concluded on the basis of a clustering
analysis of independent Canada-France Hawaii Telescope images of the field. 
The source present along the line of sight to
\afx\ exhibits [\ion{O}{2}] emission and the SED (including a Balmer
break) of a galaxy at the same redshift as the SN.  The [\ion{O}{2}] and rest-frame 
UV continuum luminosities produce similar estimates of the SFR.  These
observations are all consistent with the continuum flux from the
source along the line of sight being dominated by, if not entirely due to, 
emission from the host galaxy of \afx, and leave little room for a contribution
from a hypothetical lens galaxy at lower redshift.  These constraints
led \citet{quimbylens} to propose the existence of a dark lens from
a previously unknown population, such as a 
free-floating black hole or a dark matter halo with few baryons.

This provocative conclusion is based on the premise that \afx\ has
to be classified as a normal SN Ia and therefore the intrinsic luminosity
is known.  We demonstrated above that the spectra of \afx\ are similar to those
of both SNe Ia and Ic near maximum light (Section 4).  We also note that
 the presence of an
absorption feature from \ion{Si}{2} $\lambda$6355 does not uniquely
classify an object as a SN Ia because that line is also present in some
SNe Ic.  As an experiment, we took near-maximum light spectra of an
unambiguous nearby SNe Ic that exhibits \ion{Si}{2} features, 
SN 2004aw \citep{04aw}, resampled the data, 
and added noise to match the S/N 
of our \afx\ FIRE data that show the possible absorption feature 
near 6100~\AA.  These noisy spectra were then run through the SuperNova 
IDentification (SNID) code of \citet{snid}, using the updated templates 
of \citet{bsnip}, with care taken to exclude templates from the same object.
We found that typically all of the top 20 best matching templates are SNe Ia. 
This is in part due to the biased sample of spectral templates used by 
SN classification codes, where SNe Ia are largely overrepresented.

\citet{quimbylens} also claim that the light curve of \afx\ is a good match for
SNe Ia.  However, we note that in order to get a good reduced-$\chi^2$
for their fit, they had to artificially increase the error bars in the NIR and exclude
the $J$-band data.  Furthermore, all of the \afx\ observations prior to a phase 
of $-5$~d fall
below the SN Ia prediction, including the \zps\ non-detection at $-7.4$~d that is
$\sim$0.7~mag below their model.  This is a consequence of the fast rise time 
of \afx, which we measure above to be $11.8\pm0.7$~d if we assume that
the rise scales as $\lbol\propto t^{2}$, and $8.7\pm1.9$~d if we relax that 
constraint.  Although spectroscopically normal SNe Ia show a small
dispersion in their rise times, objects with normal light curve shapes have 
rise times of 16$-$18~d in $B$ (e.g., \citealt{hayden10,mo,gg12}).  
\citet{jha} demonstrated that the rise in $U$ (closer in rest wavelength to our 
\zps\ observations) was 2.3~d shorter, but still
significantly longer than observed for \afx.  At phases of $-12$ to $-10$~d, normal
SNe Ia emit fluxes about 20$-$40\% of the peak \citep{mo}, but that is 
about 4$\sigma$ above our non-detections (Figure~\ref{boloplot}).

In summary, although SNe Ia have reasonably similar spectra
and light curve shapes to \afx, the details of the rise time are not a good fit for 
SNe Ia and the spectra of the unambiguous SN Ic 2004aw are also similar 
to SNe Ia when degraded to our low S/N ratio.  Therefore, an identification
of \afx\ with normal SNe Ia is not required by the data, and there is no need
to invoke a new population of unseen dark gravitational lenses.  
Instead, we accept the lack of an apparent lens at face value and conclude
that \afx\ was actually an unusually luminous hydrogen-deficient SN.

\section{Conclusions}

We present multiwavelength observations of \afx\ at redshift
$z=1.388$, perhaps the most luminous SN yet discovered.  The
combination of observables presents strict constraints on any
theoretical interpretation. These are: 

\begin{itemize}

\item A peak bolometric luminosity of 4.1$\times$10$^{44}$~erg~s$^{-1}$

\item An observed rise time of $\sim$12~d

\item A fast light curve decay, with \dm = $0.95$~mag

\item At least 7$\times$10$^{50}$ erg of optical radiation emitted,
  with a total likely closer to 10$^{51}$~erg

\item Red color near maximum light ($\tbb= 6800$~K), although
  $T_{\mathrm{eff}}$ may be closer to $12,000$~K

\item Constant UV color before maximum light

\item Photospheric velocities near maximum light of
  $\sim$$11,000$~\kms

\item Spectra that most closely resemble normal SNe~Ic, with deep
  \ion{Ca}{2} P-Cygni absorption

\item Photometric and spectroscopic evidence for some line blanketing
  in the NUV from iron-peak elements, although not as much as for SNe
  Ia

\item A massive ($\sim$2$\times$10$^{10}$~\msun) host galaxy that is
  unlikely to have an extremely low metallicity

\end{itemize}

We surveyed existing models for SLSNe and found that none were
acceptable.  In particular, the large inferred \rbb\ near maximum
light is very hard to reconcile with the fast observed rise time and
the measured photospheric velocities because the SN ejecta need too
much time to reach such large radii.  The magnetar models that match
the peak luminosity and rise time do so by producing a much higher
temperature at a smaller radius.  These flaws are generic to models
powered by internal energy sources, although if the onset of the
internal energy source can be sufficiently delayed after the initial
explosion, then maybe the conflict between the velocities and radii
can be avoided. 
Shock breakout scenarios
invoking dense CSM provide a promising solution by allowing the SN to
explode well before the optical emission is detectable.  If the CSM is
sufficiently dense out to $\sim$5$\times$10$^{15}$~cm, then the light
curve will only rise after the ejecta have had time to reach that
radius.  However, 
these models leave the normal SN Ic-like spectrum unexplained and are
in apparent conflict with the observed color evolution before maximum
light.  More detailed radiative transfer calculations of the models
are necessary to know whether these flaws can be avoided.  In
addition, it is not clear how to produce the special CSM structure
that has to be assumed (a truncated wind or a shell).

\afx\ was initially identified as an object of interest due to its
unusual colors ($\rps - \ips \approx$ 2~mag at peak) and would not
have been found without the PS1 observation strategy, which includes
regular \ips\ and \zps\ observations as part of the search.  At
redshifts below $\sim$1, an object like \afx\ would have a strong
detection in \rps\ and would not stand out without knowing the
redshift.  Unlike many of 
the SN 2005ap-like SNe, the time-dilated light curve of a \afx-like SN
at lower redshift would not be unusually long and would not be
associated with the attention-grabbing lack of a visible host galaxy.
However, \afx\ was more than a magnitude brighter than its host
galaxy, which potentially provides a selection criterion to increase
the odds of finding similar objects in the future.
In the sample of SLSNe found by PS1, SN 2005ap-like objects
\citep{laura,11bam,bzj} outnumber \afx\ by
at least an order of magnitude.  Our poor understanding of
spectroscopic incompleteness and the relevant selection effects
precludes a more precise estimate of rates at this time. 

An additional complication to be considered is the role of
metallicity.  The massive host of \afx\ is in contrast with the
low-mass \citep{neill11} and low-metallicity \citep{young10,chen,bzj}
hosts of known hydrogen-poor SLSNe.  The natal composition of the
progenitor could have an indirect effect on the explosion through
stellar evolutionary processes or it could directly affect the
appearance through the opacity in the outer ejecta (could SN
2005ap-like events with higher metal abundances exhibit 
line-blanketed spectra like \afx?). 

Most of the published SLSNe that lack hydrogen in their spectra are
similar to SN 2005ap and SCP06F6
\citep{quimby05ap,barbary,quimby11,10gx,laura,06oz,bzj}. SN 2007bi
\citep{galyam09,young10} and a couple of as-yet unpublished objects
\citep{galyam12} were the only known exceptions.  They exhibited
rather different spectra and had very slow light curve decays, which
were interpreted by \citet{galyam09} to be the result of decay of a
large amount of radioactive $^{56}$Ni produced in a pair-instability
explosion.  This result has been challenged on theoretical grounds by
\citet{dessart_pi}, who show that the observations do not match
theoretical expectations for pair-instability SNe.  With the discovery
of \afx, we now have further evidence that the most luminous SNe are a
heterogeneous lot, even when just considering the objects lacking
hydrogen.  A key question for future
investigations is whether these diverse outcomes of stellar evolution
can be produced by variations on a single underlying physical model or
whether multiple pathways exist to produce SNe of these extraordinary
luminosities.  \afx\ differs from existing SLSNe-I in almost every
observable except for the peak luminosity.  It is hard to understand
how all of these differences could be produced with only small
modifications to a single model, and may indicate that \afx\ is the
first example of a different channel for producing SLSNe.

\acknowledgments
This paper includes data gathered with the 6.5-m Magellan Telescopes
located at Las Campanas Observatory, Chile.  
We thank the staffs at PS1, Gemini, Magellan, and the VLT for their
assistance with performing these observations.  We are grateful for
the grants of DD time at Gemini and the VLT. We acknowledge
J. Strader for obtaining some FourStar observations, as well as the
MMIRS observers who helped obtained some of the data here during the
fall 2010 queue observations: M. Kriek, I. Labbe, J. Roll, D. Sand,
and S. Wuyts.  SJS acknowledges funding from the European Research
Council under the European Union's Seventh Framework Programme
(FP7/2007-2013)/ERC Grant agreement n$^{\rm o}$ 291222.
The Pan-STARRS1 Surveys (PS1) have been made possible through
contributions of the Institute for Astronomy, the University of
Hawaii, the Pan-STARRS Project Office, the Max-Planck Society and its
participating institutes, the Max Planck Institute for Astronomy,
Heidelberg and the Max Planck Institute for Extraterrestrial Physics,
Garching, The Johns Hopkins University, Durham University, the
University of Edinburgh, Queen's University Belfast, the
Harvard-Smithsonian Center for Astrophysics, the Las Cumbres
Observatory Global Telescope Network Incorporated, the National
Central University of Taiwan, the Space Telescope Science Institute,
and the National Aeronautics and Space Administration under Grant
No. NNX08AR22G issued through the Planetary Science Division of the
NASA Science Mission Directorate. 
Some observations were obtained under Program IDs GN-2010B-Q-5 (PI:
Berger), GS-2010B-Q-4 (PI: Berger), and GN-2010B-DD-2 (PI: Chornock)
at the Gemini Observatory, which is operated by the Association of
Universities for  
    Research in Astronomy, Inc., under a cooperative agreement with the
    NSF on behalf of the Gemini partnership: the National Science
    Foundation (United States), the Science and Technology Facilities
    Council (United Kingdom), the National Research Council (Canada),
    CONICYT (Chile), the Australian Research Council (Australia),
    Minist\'{e}rio da Ci\^{e}ncia, Tecnologia e Inova\c{c}\~{a}o (Brazil)
    and Ministerio de Ciencia, Tecnolog\'{i}a e Innovaci\'{o}n Productiva
    (Argentina).  
Some observations were collected at the European Organisation for
Astronomical Research in the Southern Hemisphere, Chile under DDT
programme 286.D-5005 (PI : Smartt). 
The National Radio Astronomy Observatory is a facility of the National
Science Foundation operated under cooperative agreement by Associated 
Universities, Inc. 
Some of the archival data presented in this paper were taken under
programs GO-5623 and GO-9114 (PI: Kirshner) and were obtained from the
Mikulski Archive for Space Telescopes (MAST). STScI is operated by the
Association of Universities for Research in Astronomy, Inc., under
NASA contract NAS5-26555. 
Some of the computations in this paper were run on the Odyssey cluster 
supported by the FAS Science Division Research Computing Group at
Harvard University.

{\it Facilities:} \facility{PS1 (GPC1)}, \facility{Magellan:Baade
  (IMACS, FIRE, FourStar)}, \facility{Magellan:Clay (LDSS3,MMIRS)},
\facility{Gemini:Gillett (GMOS-N,NIRI)}, \facility{Gemini:South
  (GMOS-S)}, \facility{EVLA}, \facility{VLT:Yepun (HAWK-I)}


\begin{thebibliography}{}

\bibitem[Ahn et al.(2012)]{SDSS}Ahn, C.~P., Alexandroff, R., Allende
  Prieto, C., et al.\ 2012, \apjs, 203, 21  

\bibitem[Alard \& Lupton(1998)]{isis} Alard, C., \& Lupton,
  R.~H.\ 1998, \apj, 503, 325   

\bibitem[Arnett(1982)]{arnett82}Arnett, W.~D.\ 1982, \apj, 253, 785 

\bibitem[Barbary et al.(2009)]{barbary}Barbary, K., Dawson, K.~S.,
  Tokita, K., et al.\ 2009, \apj, 690, 1358

\bibitem[Barkat et al.(1967)]{barkat67}Barkat, Z., Rakavy, G., \&
  Sack, N.\ 1967, Physical Review Letters, 18, 379  

\bibitem[Berger et al.(2012)]{11bam} Berger, E., Chornock, R., Lunnan,
  R., et al.\ 2012, \apjl, 755, L29  

\bibitem[Blondin \& Tonry(2007)]{snid}Blondin, S., \& Tonry, J.~L.\ 2007, \apj, 666, 1024 

\bibitem[Branch et al.(2002)]{branch02}Branch, D., Benetti, S., Kasen,
  D., et al.\ 2002, \apj, 566, 1005  

\bibitem[Branch et al.(2000)]{branchtop}Branch, D., Jeffery, D.~J.,
  Blaylock, M., \& Hatano, K.\ 2000, \pasp, 112, 217  

\bibitem[Cardelli et al.(1989)]{ccm} Cardelli, J.~A., Clayton, G.~C.,
  \& Mathis, J.~S.\ 1989, \apj, 345, 245 

\bibitem[Chatzopoulos et al.(2012)]{manos12}Chatzopoulos, E., Wheeler,
  J.~C., \& Vinko, J.\ 2012, \apj, 746, 121  

\bibitem[Chatzopoulos et al.(2011)]{08am} Chatzopoulos, E., Wheeler,
  J.~C., Vinko, J., et al.\ 2011, \apj, 729, 143

\bibitem[Chen et al.(2013)]{chen}Chen, T.-W., Smartt, S.~J., Bresolin,
  F., et al.\ 2013, \apjl, 763, L28  

\bibitem[Chevalier(2012)]{chevalier12} Chevalier, R.~A.\ 2012, \apjl,
  752, L2  

\bibitem[Chevalier \& Irwin(2011)]{ci11} Chevalier, R.~A., \& Irwin,
  C.~M.\ 2011, \apjl, 729, L6  

\bibitem[Chomiuk et al.(2011)]{laura} Chomiuk, L., Chornock, R.,
  Soderberg, A.~M., et al.\ 2011, \apj, 743, 114  

\bibitem[Chugai(2001)]{chugai01}Chugai, N.~N.\ 2001, \mnras, 326, 1448

\bibitem[Dessart et al.(2012a)]{dessart_mag} Dessart, L., Hillier, 
D.~J., Waldman, R., Livne, E., \& Blondin, S.\ 2012, \mnras, 426, L76 

\bibitem[Dessart et al.(2012b)]{dessart_pi} Dessart, L., Waldman, 
R., Livne, E., Hillier, D.~J., \& Blondin, S.\ 2012, \mnras, 244 

\bibitem[Dexter \& Kasen(2012)]{dexter} Dexter, J., \& Kasen,
  D.\ 2012, arXiv:1210.7240 

\bibitem[Drake et al.(2011)]{drakesn}Drake, A.~J., Djorgovski, S.~G.,
  Mahabal, A., et al.\ 2011, \apj, 735, 106 

\bibitem[Drake et al.(2010)]{drake08fz} Drake, A.~J., Djorgovski, 
S.~G., Prieto, J.~L., et al.\ 2010, \apjl, 718, L127 

\bibitem[Dressler et al.(2006)]{imacspaper} Dressler, A., Hare, 
T., Bigelow, B.~C., \& Osip, D.~J.\ 2006, \procspie, 6269,  

\bibitem[Drout et al.(2011)]{drout11} Drout, M.~R., Soderberg, 
A.~M., Gal-Yam, A., et al.\ 2011, \apj, 741, 97 

\bibitem[Erb et al.(2012)]{erb12} Erb, D.~K., Quider, A.~M., Henry,
  A.~L., \& Martin, C.~L.\ 2012, \apj, 759, 26  

\bibitem[Faber et al.(2007)]{faber07}Faber, S.~M., Willmer, C.~N.~A.,
  Wolf, C., et al.\ 2007, \apj, 665, 265  

\bibitem[{{Filippenko}(1982)}]{fil82} {Filippenko}, A.~V. 1982, PASP,
  94, 715

\bibitem[Foley(2013)]{foleyuv}Foley, R.~J. 2013, \mnras, submitted,
  arXiv:1212.6261

\bibitem[Foley et al.(2012)]{foley11iv}Foley, R.~J., Kromer, M.,
  Marion, G., et al.\ 2012, \apjl, 753, L5

\bibitem[Foley et al.(2003)]{foley02ap}Foley, R.~J., Papenkova, 
M.~S., Swift, B.~J., et al.\ 2003, \pasp, 115, 1220 

\bibitem[Fransson et al.(2005)]{fransson98s}Fransson, C., Challis,
  P.~M., Chevalier, R.~A., et al.\ 2005, \apj, 622, 991  

\bibitem[Fynbo et al.(2009)]{fynbo09} Fynbo, J.~P.~U., Jakobsson, P.,
  Prochaska, J.~X., et al.\ 2009, \apjs, 185, 526  

\bibitem[Gabasch et al.(2004)]{gabasch04} Gabasch, A., Bender, R.,
  Seitz, S., et al.\ 2004, \aap, 421, 41 

\bibitem[Gal-Yam(2012)]{galyam12}Gal-Yam, A.\ 2012, Science, 337, 927

\bibitem[Gal-Yam et al.(2009)]{galyam09}Gal-Yam, A., Mazzali, P.,
  Ofek, E.~O., et al.\ 2009, \nat, 462, 624  

\bibitem[Galama et al.(1998)]{galama}Galama, T.~J., Vreeswijk, P.~M.,
  van Paradijs, J., et al.\ 1998, \nat, 395, 670

\bibitem[Ganeshalingam et al.(2011)]{mo} Ganeshalingam, 
M., Li, W., \& Filippenko, A.~V.\ 2011, \mnras, 416, 2607 

\bibitem[Gezari et al.(2009)]{gezari08es}Gezari, S., Halpern, J.~P.,
  Grupe, D., et al.\ 2009, \apj, 690, 1313  

\bibitem[Ginzburg \& Balberg(2012)]{gb12} Ginzburg, S., \& Balberg,
  S.\ 2012, \apj, 757, 178  

\bibitem[Gonz{\'a}lez-Gait{\'a}n et al.(2012)]{gg12} 
Gonz{\'a}lez-Gait{\'a}n, S., Conley, A., Bianco, F.~B., et al.\ 2012, \apj, 
745, 44 

\bibitem[Greisen(2003)]{aips} Greisen, E.~W.\ 2003, 
Information Handling in Astronomy - Historical Vistas, 285, 109 

\bibitem[Hatano et al.(1999)]{hatano99}Hatano, K., Branch, D.,
  Fisher, A., Millard, J., \& Baron, E.\ 1999, \apjs, 121, 233  

\bibitem[Hayden et al.(2010)]{hayden10} Hayden, B.~T., 
Garnavich, P.~M., Kessler, R., et al.\ 2010, \apj, 712, 350 

\bibitem[Hilbert et al.(2008)]{hilbert08} Hilbert, S., White,  S.~D.~M., 
Hartlap, J., \& Schneider, P.\ 2008, \mnras, 386, 1845 

\bibitem[Hjorth et al.(2012)]{hjorth12}Hjorth, J., Malesani, D.,
  Jakobsson, P., et al.\ 2012, \apj, 756, 187  

\bibitem[Hodapp et al.(2003)]{niri} Hodapp, K.~W., Jensen, J.~B.,
  Irwin, E.~M., et al.\ 2003, \pasp, 115, 1388  

\bibitem[Hodapp et al.(2004)]{PS1_optics} Hodapp, K.~W., Siegmund, 
W.~A., Kaiser, N., Chambers, K.~C., Laux, U., Morgan, J., 
\& Mannery, E.\ 2004, \procspie, 5489, 667 

\bibitem[H{\"o}flich et al.(1999)]{hoflich99}H{\"o}flich, P., Wheeler,
  J.~C., \& Wang, L.\ 1999, \apj, 521, 179

\bibitem[Hook et al.(2004)]{hookgmos} Hook, I.~M., J{\o}rgensen, 
I., Allington-Smith, J.~R., et al.\ 2004, \pasp, 116, 425 

\bibitem[Iwamoto et al.(1994)]{iwamoto94i}Iwamoto, K., Nomoto, K.,
  H\"{o}flich, P., et al.\ 1994, \apjl, 437, L115  

\bibitem[Jeffery et al.(1994)]{jeffery93j} Jeffery, D.~J., 
Kirshner, R.~P., Challis, P.~M., et al.\ 1994, \apjl, 421, L27 

\bibitem[Jha et al.(2006)]{jha} Jha, S., Kirshner, R.~P., 
Challis, P., et al.\ 2006, \aj, 131, 527 

\bibitem[Kaiser et al.(2010)]{PS1_system} Kaiser, N., et al. \ 2010,
  \procspie, 7733,  12K  

\bibitem[Kajisawa et al.(2010)]{moircs}Kajisawa, M., Ichikawa, T.,
  Yamada, T., et al.\ 2010, \apj, 723, 129 

\bibitem[Kasen \& Bildsten(2010)]{kb10} Kasen, D., \& Bildsten,
  L.\ 2010, \apj, 717, 245   

\bibitem[Kennicutt(1998)]{kennicutt98} Kennicutt, R.~C., Jr.\ 1998,
  \araa, 36, 189 

\bibitem[Kirshner et al.(1993)]{kirshner93}Kirshner, R.~P., Jeffery,
  D.~J., Leibundgut, B., et al.\ 1993, \apj, 415, 589  

\bibitem[Kissler-Patig et al.(2008)]{hawki}
  Kissler-Patig, M., Pirard, J.-F., Casali, M., et al.\ 2008, \aap,
  491, 941  

\bibitem[Komatsu et al.(2011)]{WMAP} Komatsu, E., Smith, 
K.~M., Dunkley, J., et al.\ 2011, \apjs, 192, 18 

\bibitem[Leloudas et al.(2012)]{06oz} Leloudas, G., Chatzopoulos, E.,
  Dilday, B., et al.\ 2012, \aap, 541, A129 

\bibitem[Lunnan et al.(2013)]{bzj}Lunnan, R., Chornock, R., Berger,
  E., et al. 2013, ApJ, submitted, arXiv:1303.1531

\bibitem[McLeod et al.(2012)]{mmirs}McLeod, B., Fabricant, D.,
  Nystrom, G., et al.\ 2012, \pasp, 124, 1318  

\bibitem[Maeda et al.(2006)]{maeda06}Maeda, K., Mazzali, P.~A., \&
  Nomoto, K.\ 2006, \apj, 645, 1331

\bibitem[Maeda et al.(2007)]{maeda05bf} Maeda, K., Tanaka, M., Nomoto,
  K., et al.\ 2007, \apj, 666, 1069  

\bibitem[Magnier(2006)]{PS1_IPP} Magnier, E.\ 2006, Proceedings of The Advanced 
Maui Optical and Space Surveillance Technologies Conference, Ed.:
S. Ryan, The Maui Economic Development Board, p.E5 

\bibitem[Maguire et al.(2012)]{maguire11fe}Maguire, K., Sullivan, M.,
  Ellis, R.~S., et al.\ 2012, \mnras, 426, 2359 

\bibitem[Mannucci et al.(2010)]{fmr}Mannucci, F., Cresci, G.,
  Maiolino, R., Marconi, A., \& Gnerucci, A.\ 2010, \mnras, 408, 2115 

\bibitem[Maraston(2005)]{maraston05} Maraston, C.\ 2005, \mnras, 362,
  799 

\bibitem[Mazzali et al.(2002)]{mazzali02ap}Mazzali, P.~A., Deng, J.,
  Maeda, K., et al.\ 2002, \apjl, 572, L61  

\bibitem[Metzger(2010)]{metzger10}Metzger, B.~D.\ 2010, \mnras, 409,
  284

\bibitem[Millard et al.(1999)]{millard99}Millard, J., Branch, D.,
  Baron, E., et al.\ 1999, \apj, 527, 746 

\bibitem[Miller et al.(2009)]{miller08es}Miller, A.~A., Chornock, R.,
  Perley, D.~A., et al.\ 2009, \apj, 690, 1303  

\bibitem[Moriya \& Maeda(2012)]{mm12} Moriya, T.~J., \& Maeda,
  K.\ 2012, \apjl, 756, L22 

\bibitem[Moriya \& Tominaga(2012)]{mt12} Moriya, T.~J., \& Tominaga,
  N.\ 2012, \apj, 747, 118  

\bibitem[Moriya et al.(2013)]{moriya06gy} Moriya, T.~J., Blinnikov,
  S.~I., Tominaga, N., et al.\ 2013, \mnras, 428, 1020  

\bibitem[Nakar \& Sari(2010)]{nakar} Nakar, E., \& Sari, R.\ 2010,
  \apj, 725, 904  

\bibitem[Neill et al.(2011)]{neill11}Neill, J.~D., Sullivan, M.,
  Gal-Yam, A., et al.\ 2011, \apj, 727, 15  

\bibitem[Nomoto et al.(1993)]{nomoto93j}Nomoto, K., Suzuki, T.,
  Shigeyama, T., et al.\ 1993, \nat, 364, 507  

\bibitem[Nugent et al.(2011)]{nugent11fe}Nugent, P.~E., Sullivan, M.,
  Cenko, S.~B., et al.\ 2011, \nat, 480, 344  

\bibitem[Ofek et al.(2007)]{ofek06gy}Ofek, E.~O., Cameron, P.~B.,
  Kasliwal, M.~M., et al.\ 2007, \apjl, 659, L13 

\bibitem[Ofek et al.(2010)]{ofek09uj}Ofek, E.~O., Rabinak, I., 
Neill, J.~D., et al.\ 2010, \apj, 724, 1396 

\bibitem[Pastorello et al.(2010)]{10gx} Pastorello, A., 
Smartt, S.~J., Botticella, M.~T., et al.\ 2010, \apjl, 724, L16 

\bibitem[Perley et al.(2013)]{perley}Perley, D.~A., Levan, A.~J.,
  Tanvir, N.~R., et al.\ 2013, ApJ, submitted, arXiv:1301.5903  

\bibitem[Perley et al.(2011)]{evla}Perley, R.~A., Chandler, 
C.~J., Butler, B.~J., \& Wrobel, J.~M.\ 2011, \apjl, 739, L1 

\bibitem[Persson et al.(2008)]{fourstar} Persson, S.~E., 
Barkhouser, R., Birk, C., et al.\ 2008, \procspie, 7014, 95P 

\bibitem[Podsiadlowski et al.(1993)]{pods93j}Podsiadlowski, P., Hsu,
  J.~J.~L., Joss, P.~C., \& Ross, R.~R.\ 1993, \nat, 364, 509  

\bibitem[Quataert \& Shiode(2012)]{qs12} Quataert, E., \& Shiode,
  J.\ 2012, \mnras, 423, L92  

\bibitem[Quimby et al.(2007)]{quimby05ap}Quimby, R.~M., Aldering, G.,
  Wheeler, J.~C., et al.\ 2007, \apjl, 668, L99  

\bibitem[Quimby et al.(2011)]{quimby11} Quimby, R.~M., Kulkarni, 
S.~R., Kasliwal, M.~M., et al.\ 2011, \nat, 474, 487 

\bibitem[Quimby et al.(2013)]{quimbylens} Quimby, R.~M., Werner, 
M.~C., Oguri, M., et al.\ 2013, arXiv:1302.2785 

\bibitem[Rakavy \& Shaviv(1967)]{rakavy67}Rakavy, G., \& Shaviv,
  G.\ 1967, \apj, 148, 803  

\bibitem[Rest et al.(2011)]{rest03ma}Rest, A., Foley, R.~J., Gezari,
  S., et al.\ 2011, \apj, 729, 88  

\bibitem[Rest et al.(2005)]{rest05} Rest, A., Stubbs, C., Becker,
  A.~C., et al.\ 2005, \apj, 634, 1103  

\bibitem[Richmond et al.(1994)]{richmond94} Richmond, M.~W., 
Treffers, R.~R., Filippenko, A.~V., et al.\ 1994, \aj, 107, 1022 

\bibitem[Richmond et al.(1996)]{richmond96} Richmond, M.~W., van 
Dyk, S.~D., Ho, W., et al.\ 1996, \aj, 111, 327 

\bibitem[Riess et al.(2011)]{shoes} Riess, A.~G., Macri, L., 
Casertano, S., et al.\ 2011, \apj, 730, 119 

\bibitem[Riess et al.(2007)]{riess07} Riess, A.~G., Strolger, 
L.-G., Casertano, S., et al.\ 2007, \apj, 659, 98 

\bibitem[Sauer et al.(2008)]{sauer08}Sauer, D.~N., Mazzali, 
P.~A., Blondin, S., et al.\ 2008, \mnras, 391, 1605 

\bibitem[Sauer et al.(2006)]{sauer06} Sauer, D.~N., Mazzali, 
P.~A., Deng, J., et al.\ 2006, \mnras, 369, 1939 

\bibitem[Schechter(1976)]{schechter}Schechter, P.\ 1976, \apj, 203, 297 

\bibitem[Schlafly \& Finkbeiner(2011)]{eddiedoug} Schlafly, E.~F., \&
  Finkbeiner, D.~P.\ 2011, \apj, 737, 103  

\bibitem[Schlegel et al.(1998)]{sfd98}Schlegel, D.~J., Finkbeiner,
  D.~P., \& Davis, M.\ 1998, \apj, 500, 525 

\bibitem[Silverman et al.(2012)]{bsnip} Silverman, J.~M., 
Foley, R.~J., Filippenko, A.~V., et al.\ 2012, \mnras, 425, 1789 

\bibitem[Simcoe et al.(2008)]{fireref}Simcoe, R.~A., et al.\ 2008,
  \procspie, 7014, 27

\bibitem[Smith et al.(2008)]{smith06tf}Smith, N., Chornock, R., Li,
  W., et al.\ 2008, \apj, 686, 467  

\bibitem[Smith et al.(2010)]{smith06gyb}Smith, N., Chornock, R.,
  Silverman, J.~M., Filippenko, A.~V., \& Foley, R.~J.\ 2010, \apj,
  709, 856  

\bibitem[Smith et al.(2007)]{smith06gya}Smith, N., Li, W., Foley,
  R.~J., et al.\ 2007, \apj, 666, 1116 

\bibitem[Smith \& McCray(2007)]{smithmccray} Smith, N., \& McCray,
  R.\ 2007, \apjl, 671, L17  

\bibitem[Stanek et al.(2006)]{stanek06}Stanek, K.~Z., Gnedin, 
O.~Y., Beacom, J.~F., et al.\ 2006, Acta Astronomica, 56, 333 

\bibitem[Stritzinger et al.(2006)]{max06} Stritzinger, M., Mazzali,
  P.~A., Sollerman, J., \& Benetti, S.\ 2006, \aap, 460, 793  

\bibitem[Stubbs et al.(2010)]{PS_lasercal} Stubbs, C.~W., Doherty, 
P., Cramer, C., Narayan, G., Brown, Y.~J., Lykke, K.~R., Woodward, J.~T., 
\& Tonry, J.~L.\ 2010, \apjs, 191, 376 

\bibitem[Takahashi et al.(2011)]{taka11} Takahashi, R., Oguri, 
M., Sato, M., \& Hamana, T.\ 2011, \apj, 742, 15 

\bibitem[Taubenberger et al.(2006)]{04aw} Taubenberger, S., 
Pastorello, A., Mazzali, P.~A., et al.\ 2006, \mnras, 371, 1459 

\bibitem[Thomas et al.(2011)]{thomas11}Thomas, R.~C., Nugent, P.~E.,
  \& Meza, J.~C.\ 2011, \pasp, 123, 237  

\bibitem[Tonry \& Onaka(2009)]{PS1_GPCA} Tonry, J., \& Onaka,
  P.\ 2009, Advanced Maui Optical and Space Surveillance Technologies
  Conference,  Proceedings of the Advanced Maui Optical and Space
  Surveillance Technologies Conference, Ed.: S. Ryan, p.E40.    

\bibitem[Tonry et al.(2012)]{JTphoto} Tonry, J.~L., Stubbs, 
C.~W., Lykke, K.~R., et al.\ 2012, \apj, 750, 99 

\bibitem[Vacca et al.(2003)]{vacca03} Vacca, W.~D., Cushing, 
M.~C., \& Rayner, J.~T.\ 2003, \pasp, 115, 389 

\bibitem[Valenti et al.(2008)]{03jd} Valenti, S., Benetti, 
S., Cappellaro, E., et al.\ 2008, \mnras, 383, 1485 

\bibitem[Wang \& Wheeler(2008)]{ww08}Wang, L., \& Wheeler,
  J.~C.\ 2008, \araa, 46, 433  

\bibitem[Wheeler et al.(1994)]{wheeler94}Wheeler, J.~C., Harkness,
  R.~P., Clocchiatti, A., et al.\ 1994, \apjl, 436, L135  

\bibitem[Woosley(2010)]{woosley10} Woosley, S.~E.\ 2010, \apjl, 719,
  L204 

\bibitem[Woosley et al.(2007)]{woosley07}Woosley, S.~E., Blinnikov,
  S., \& Heger, A.\ 2007, \nat, 450, 390  

\bibitem[Woosley et al.(1994)]{woosley93j}Woosley, S.~E., Eastman,
  R.~G., Weaver, T.~A., \& Pinto, P.~A.\ 1994, \apj, 429, 300  

\bibitem[Young et al.(2010)]{young10}Young, D.~R., Smartt, S.~J.,
  Valenti, S., et al.\ 2010, \aap, 512, A70  

\end{thebibliography}
\end{document}